\documentclass[apj]{emulateapj}
\usepackage{textcomp}
\usepackage{epsfig, epstopdf}
\usepackage{longtable}
\usepackage{lscape}
\usepackage{amssymb,amsmath}

\newcommand{\cf}{{\ifmmode{C_{\rm f}}\else{$C_{\rm f}$}\fi}}
\newcommand{\cfov}{{\ifmmode{C_{\rm fov}}\else{$C_{\rm fov}$}\fi}}
\newcommand{\cfratio}{{\ifmmode{C_{\rm fov\_ratio}}\else{$C_{\rm
        fov\_ratio}$}\fi}}
\newcommand{\cfnorm}{{\ifmmode{C_{\rm fov\_norm}}\else{$C_{\rm
        fov\_norm}$}\fi}}
\newcommand{\zem}{{\ifmmode{z_{\rm em}}\else{$z_{\rm em}$}\fi}}
\newcommand{\zabs}{{\ifmmode{z_{\rm abs}}\else{$z_{\rm abs}$}\fi}}
\newcommand{\kms}{{\ifmmode{{\rm km~s}^{-1}}\else{km~s$^{-1}$}\fi}}
\newcommand{\voff}{{\ifmmode{v_{\rm off}}\else{$v_{\rm off}$}\fi}}
\newcommand{\vrot}{{\ifmmode{v_{\rm rot}}\else{$v_{\rm rot}$}\fi}}
\newcommand{\cm}{{\ifmmode{{\rm cm}^{-1}}\else{cm$^{-1}$}\fi}}
\newcommand{\cmm}{{\ifmmode{{\rm cm}^{-2}}\else{cm$^{-2}$}\fi}}
\newcommand{\cmmm}{{\ifmmode{{\rm cm}^{-3}}\else{cm$^{-3}$}\fi}}
\newcommand{\lya}{Ly$\alpha$} 
\newcommand{\logN}{{\ifmmode{{\rm log}N}\else{log$N$}\fi}}

\newcounter{species} 
\def\ion#1#2{\setcounter{species}{#2}#1$\;${\scriptsize\Roman{species}}\relax}

\shorttitle{MCMC Fitting to a Mini-BAL in UM675}
\shortauthors{Ishita et al.}

\begin{document}

\title{MCMC-based Voigt Profile fitting to a Mini-BAL System in the
  Quasar UM675\altaffilmark{1,2}}

\footnotetext[1]{Data presented herein were obtained at the W.M. Keck
  Observatory, which is operated as a scientific partnership among the
  California Institute of Technology, the University of California and
  the National Aeronautics and Space Administration. The Observatory
  was made possible by the generous financial support of the W.M. Keck
  Foundation.}
\footnotetext[2]{Based on data collected at Subaru Telescope, which is
  operated by the National Astronomical Observatory of Japan.}

\author{Dai Ishita}
\affiliation{Department of Physics, Faculty of Science, Shinshu
  University, 3-1-1 Asahi, Matsumoto, Nagano 390-8621, Japan}
\author{Toru Misawa}
\affil{School of General Education, Shinshu University, 3-1-1 Asahi,
  Matsumoto, Nagano 390-8621, Japan}
\author{Daisuke Itoh}
\affiliation{Department of Physics, Faculty of Science, Shinshu
  University, 3-1-1 Asahi, Matsumoto, Nagano 390-8621, Japan}
\author{Jane C. Charlton} \affiliation{Department
  of Astronomy \& Astrophysics, Pennsylvania State University, 525
  Davey Lab, University Park, PA 16802}
\author{Michael Eracleous}
\affiliation{Department of Astronomy \& Astrophysics, Pennsylvania
  State University, 525 Davey Lab, University Park, PA 16802}
\affiliation{Institute for Gravitation and the Cosmos, The
  Pennsylvania State University, University Park, PA 16802}

\begin{abstract}
We introduce a Bayesian approach coupled with a Markov Chain Monte
Carlo (MCMC) method and the maximum likelihood statistic for fitting
the profiles of narrow absorption lines (NALs) in quasar spectra.
This method also incorporates overlap between different absorbers. We
illustrate and test this method by fitting models to a ``mini-broad''
(mini-BAL) and six NAL profiles in four spectra of the quasar UM675
taken over a rest-frame interval of 4.24~years.  Our fitting results
are consistent with past results for the mini-BAL system in this
quasar by \citet{ham97b}. We also measure covering factors (\cf) for
two narrow components in the \ion{C}{4} and \ion{N}{5} mini-BALs and
their overlap covering factor with the broad component.  We find that
\cf~(\ion{N}{5}) is always larger than \cf~(\ion{C}{4}) for the broad
component, while the opposite is true for the narrow components in the
mini-BAL system.  This could be explained if the broad and narrow
components originated in gas at different radial distances, but it
seems more likely to be due to them produced by gas at the same
distance but with different gas densities (i.e., ionization states).
The variability detected only in the broad absorption component in the
mini-BAL system is probably due to gas motion since both
\cf~(\ion{C}{4}) and \cf~(\ion{N}{5}) vary.  We determine for the
first time that multiple absorbing clouds (i.e., a broad and two
narrow components) overlap along our line of sight. We conclude that
the new method improves fitting results considerably compared to
previous methods.
\end{abstract}


\section{Introduction} \label{sec:intro}
Astronomical objects along our line of sight to distant quasars will
produce absorption lines in the spectra of those quasars (hereafter,
quasar absorption lines or QALs). The absorbers include not only
cosmologically intervening objects, like foreground galaxies and
intergalactic medium (IGM) (hereafter, {\it intervening} QALs), but
also gas clouds that are physically associated with the quasars
themselves such as AGN outflows (hereafter, {\it intrinsic}
QALs). These absorption lines have historically been studied by
counting their numbers as a function of redshift ($dN/dz$) for various
transitions to place constraints on their physical sizes and comoving
number densities (e.g., \citealt{ste90,bec94,lan95}).  With the advent
of 8--10~m class telescopes and their high-dispersion spectrographs
such as the Keck telescope with the High Resolution Echelle
Spectrometer (HIRES), the Very Large Telescope (VLT) with the
Ultraviolet and Visual Echelle Spectrograph (UVES), and the Subaru
telescope with the High Dispersion Spectrograph (HDS), we are now able
to measure physical parameters in addition to the absorption redshift
(\zabs), e.g., the column density ($N$), and the Doppler parameter
($b$) by fitting Voigt profiles (VPs) to the absorption lines.

The intrinsic QALs are generally blueshifted\footnote{In rare cases
  they are redshifted by up to a few thousand
  \kms~\citep{hal02,mis07a}.}  from the quasar emission redshift by up
to $\sim$0.2--0.3~c \citep{ham18}, which suggests that they are
accelerated away from the quasar by several possible mechanisms,
including radiative pressure \citep{mur95,pro00}, magnetocentrifugal
force \citep[e.g.,][]{eve05}, and thermal pressure
\citep[e.g.,][]{che05}. They are usually classified into three
categories according to their line widths: broad absorption lines
(BALs) with total FWHM $\geq$ 2,000~km/s, narrow absorption lines
(NALs) with FWHM $\leq$ 500~km/s, and an intermediate subclass
(mini-BALs).  The difference in line width is often explained by the
inclination angle of our line of sight relative to the direction of
motion of the wind \citep[e.g.,][]{elv00,gan01,ito20} or the stage of
quasar evolution \citep[e.g.,][]{far07}, although it may also depend
on the emission line outflow properties and/or the hardness of the
ionizing spectral energy distribution (e.g., \citealt{ran20}).
Intrinsic QALs may play an important role in providing energy and
momentum feedback to the interstellar medium (ISM) and circumgalactic
medium (CGM) of their host galaxies and the surrounding IGM, and may
play a role in regulating star formation activity in their host
galaxies \citep[e.g.,][]{spr05}.

In past studies of QALs, several $\chi^2$-based VP fitting codes have
been used, including {\tt VPFIT} \citep{car91}, {\tt autoVP}
\citep{dav97}, and {\tt MINFIT} \citep{chu97,chu03}.  Recently,
\citet{lia17} employed a Bayesian approach for VP fitting with a
Markov Chain Monte Carlo (MCMC) method ({\tt BayesVP}) because the
traditional $\chi^2$-based fitting codes have several weaknesses: for
example, (a) they cannot place strong constraints on the parameters of
undetected or saturated QALs, (b) they can provide incorrect results
depending on the initial conditions if parameters have a multi-modal
probability distribution, (c) grid search is computationally expensive
when the number of fitting parameters is large.  Using {\tt BayesVP},
\citet{lia18} successfully determined or constrained the physical
parameters of absorption lines detected at low significance as well as
non-detected or saturated ones.  \citet{sam21} also introduced
Bayesian methods for fitting photoionization models to intervening
absorbers and were able to place stringent constraints on their
physical parameters.

The Bayesian approach with MCMC methods has another powerful property:
it allows us to set upper and/or lower limits in advance for each
fitting parameter as needed.  When we apply VP fits to intrinsic QALs,
we always need a fourth parameter, the covering factor (\cf; the
fraction of the flux from background sources that passes through a
foreground absorber along our line of sight; \citealt{bar95}) in
addition to the other three (\zabs, \logN, and $b$) because the size
of the corresponding absorber in the vicinity of the flux source can
be smaller than the source itself (i.e., the continuum source or broad
emission line region; BELR). By its definition, \cf\ should be between
0 and 1.

One of the $\chi^2$-based VP fit codes, {\tt MINFIT}, enables us to
perform VP fits to intrinsic QALs with \cf\ as a fourth fit parameter,
however, it sometimes gives unphysical values such as \cf\ $<$ 0 or
\cf\ $>$ 1.  \citet{mis05} showed that the measurement of the
\cf\ value is very sensitive to continuum level errors, especially for
very weak absorption lines whose real \cf\ values are close to 1. As
long as the \cf\ value is unphysical, the other fit parameters (i.e.,
\zabs, \logN, and $b$) have no physical meaning. In previous studies,
this problem was handled by assuming \cf\ = 1 for those components and
refitting to solve for their \zabs, \logN, and $b$ values, but this
procedure does not adequately cover all physical possibilities.

In this study, we employ a Bayesian approach with MCMC-based method
for fitting VPs to {\it intrinsic} QALs as \citet{lia17} did for {\it
  intervening} QALs. In Section~2, we introduce the MCMC method,
describe the fitting procedure in detail, and compare it to the
traditional $\chi^2$-based method.
As a test case, we apply the method to the \ion{C}{4} and \ion{N}{5}
mini-BALs and intervening \ion{C}{4} NALs in the spectrum of the
quasar UM675 in Section~3, and present our fitting results in
Section~4. We summarize our conclusions in Section~5. We use a
cosmology with $H_{0}$=69.6 \kms~Mpc$^{-1}$, $\Omega_{m}$=0.29, and
$\Omega_{\Lambda}$=0.71 throughout the paper.

\section{Method: MCMC-based VP Fitting} \label{sec:method}

We constrain the posterior distribution of the fit parameters,
$p$($\vec{\theta}|data$) in the Bayesian approach,
\begin{equation}
  p(\vec{\theta}|data) \propto
  \mathcal{L}(data|\vec{\theta})p(\vec{\theta}),
\end{equation}
where $\vec{\theta}$ = \{$z, \log{N}, b, C_{\rm f}$\} is the vector of
fit parameters, $p$($\vec{\theta}$) is the prior distribution, and
$\mathcal{L}$($data|\vec{\theta}$) is the total likelihood.  The
function $\mathcal{L}$($data|\vec{\theta}$) is expressed as the
product of the Gaussian likelihood for each pixel by
\begin{equation}
\mathcal{L}(data|\vec{\theta}) = \prod_{i=1}^{n} l_{i} =
\prod_{i=1}^{n} \frac{1}{\sqrt{2\pi
    \sigma^2_i}}\exp\left(-\frac{(f_{{\rm obs}\_i}-f_{{\rm
    mod}\_i})^2}{2\sigma_i^2}\right),
\end{equation}
where $l_{i}$ is the likelihood of the individual pixel, and $f_{{\rm
    obs}\_i}$, $f_{{\rm mod}\_i}$, and $\sigma_i$ are the observed
flux, the model flux, and the uncertainty of the observed flux in the
$i$-th pixel of the spectrum.

\subsection{Overlap Covering Factor} \label{sec:cfov}
\begin{figure*}
  \vspace{2cm}
  \begin{center}   
    \includegraphics[width=17cm,angle=0]{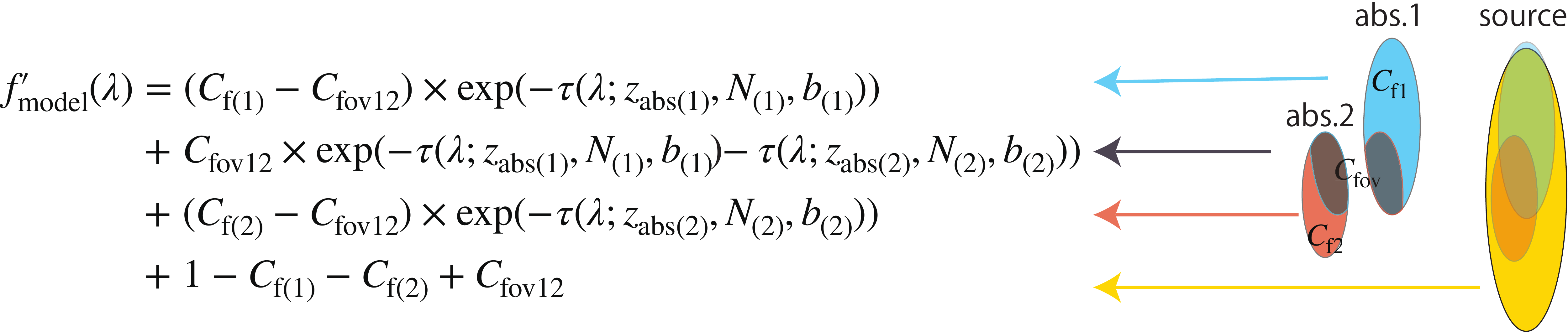}
  \end{center}
  \caption{Sketch of an absorption system with two components.  If
    multiple clouds overlap along our line of sight to the background
    source, flux from the source is absorbed more than once. The
    number of parameters needed to fit the spectrum is
    $\left[\sum_{i=2}^{n} {n \choose i} \right] + 4n$, and increases
    quickly with the number of components $n$.\label{fig:overlap}}
\end{figure*}

In addition to the physical parameters noted above, we also add the
overlap covering factor (\cfov) that is the covering factor by
multiple absorbers simultaneously along our line of sight as
illustrated in Figure~\ref{fig:overlap}. For example, if we detect two
absorbing components 1 and 2 with covering factors of \cf$_{(1)}$ =
0.4 and \cf$_{(2)}$ = 0.8 in a single system, the maximum and minimum
values of \cfov\ are max(\cfov) [$\equiv$ min(\cf$_{(1)}$,
  \cf$_{(2)}$)] = 0.4 and min(\cfov) [$\equiv$ max(0,
  \cf$_{(1)}+$\cf$_{(2)}-1$)] = 0.2, respectively. Thus, the permitted
range of \cfov\ depends on the \cf\ of each component.  To avoid any
possible biases for the overlap covering factors, we assume a uniform
distribution for the normalized covering factor
\begin{equation}
   \cfnorm = \frac{C_{\rm fov} - {\rm min}(C_{\rm fov})}{{\rm
       max}(C_{\rm fov})-{\rm min}(C_{\rm fov})}\label{eqn:cfnorm},
\end{equation}
where min(\cfov) and max(\cfov) are determined ahead of time as
described in Section~2.2.  Using these boundary values, we also
introduce the overlap covering factor ratio (\cfratio) defined by
\begin{equation}
  C_{\rm fov\_ratio} = \frac{C_{\rm fov}}{\rm max(C_{\rm
      fov})}\label{eqn:cfratio}.
\end{equation}

The traditional $\chi^2$-based {\tt MINFIT} fits models to the
observed spectra by
\begin{eqnarray}
  f_{\rm mod}(\lambda)
  & = & \prod_{j=1}^{n} \left[ C_{{\rm f}(j)} \times
  \exp{\{-\tau(\lambda; z_{{\rm abs}(j)},N_{(j)},b_{(j)})\}} \right. \nonumber \\
  & & \left. + 1 - C_{{\rm f}(j)} \right] \label{eqn:minfit},
\end{eqnarray}
where $C_{{\rm f}(j)}$, $z_{{\rm abs}(j)}$, $N_{(j)}$, and $b_{(j)}$
are the parameters of component $j$. In equation~(5), the overlap
covering factor \cfov\ is not considered, but instead the brightness
distribution of the background source, as seen by all absorbers along
the cylinder of sight, is always uniform and the same, regardless of
whether the absorbers overlap.  This assumption is incorrect except
for intervening absorbers where \cf\ = 1. Once we introduce \cfov, the
residual flux after being attenuated by two absorbing components for
example would be written as
\begin{eqnarray}
  f_{\rm mod}^{\prime}(\lambda)
  & = &  (C_{{\rm f}(1)} - C_{{\rm
      fov}(12)})  \nonumber \\
  & & \hspace{0.5cm} \times \exp{(-\tau(\lambda; z_{{\rm
        abs}(1)},N_{(1)},b_{(1)}))} \nonumber \\
  & & + (C_{{\rm f}(2)} - C_{{\rm fov}(12)})  \nonumber \\
  & & \hspace{0.5cm} \times \exp{(-\tau(\lambda;
    z_{{\rm abs}(2)},N_{(2)},b_{(2)}))} \nonumber \\
  & & + C_{{\rm fov}(12)} \times \exp\{(-\tau(\lambda; z_{{\rm
        abs}(1)},N_{(1)},b_{(1)}) \nonumber \\
  & & \hspace{0.5cm} - \tau(\lambda; z_{{\rm
        abs}(2)},N_{(2)},b_{(2)}))\} \nonumber \\
  & & + 1 - C_{{\rm f}(1)} - C_{{\rm f}(2)} + C_{{\rm fov}(12)},
\end{eqnarray}
where $C_{{\rm fov}(12)}$ is the overlap covering factor between
components 1 and 2.  Since our code can accept any number of
components, the total number of fit parameters for $n$ components is
$\left[\sum_{i=2}^{n} {n \choose i} \right] + 4n$ (where ${n \choose
  i}$ represents the number of possible combinations of $i$ items
drawn from a set of $n$ items) after adding the overlap covering
factors (the first term) to the original parameters for each component
(the second term). However, it should be noted that total number of
model parameters (and then the posterior distribution of them) would
be quite large if we use a large number of components in the fit.

\subsection{Fitting Procedure} \label{sec:proc}
Before using our MCMC-based VP fitting code ({\tt mc2fit}, hereafter),
we need the prior distribution of model parameters. Since no information
is available in advance, we assume a uniform distribution between
appropriate lower and upper limits of each parameter as the prior.  We
then apply {\tt mc2fit} to absorption lines following the procedure
below:

\begin{itemize}
\item{We make an initial estimate of the values of the parameters for
  each absorption component (i.e., \zabs, \logN, $b$, and \cf) except
  for the overlap covering factor \cfov, using the $\chi^2$-based VP
  fit code {\tt MINFIT}.  The code automatically returns the minimum
  number of necessary components for fitting the absorption profile.
  From these we determine min(\cfov) and max(\cfov), used in
  equation~(3).}
\item{We start the MCMC sampling using the affine-invariant ensemble
  sampler proposed by \citet{goo10} in the model parameter space with
  a number of dimensions of $\left[\sum_{i=2}^{n} {n \choose i}
    \right] + 4n$.  When the sampling points (i.e.,
  walkers\footnote{Discrete points that move around to sample the
    $N$-dimensional parameter space.}; the default number is 256 in
  our calculations) move from the position of the roughly estimated
  values, we demand they should fulfill detailed
  balance\footnote{Moving process of walkers should be in balance with
    its inverse process as expressed by
    $f$($\vec{\theta_2}|\vec{\theta_1}$)$\times$$f$($\vec{\theta_1}$)
    =
    $f$($\vec{\theta_1}|\vec{\theta_2}$)$\times$$f$($\vec{\theta_2}$),
    where $f$($\vec{\theta_i}$) is the posterior probability density
    at $\vec{\theta_1}$ and $f$($\vec{\theta_j}|\vec{\theta_i}$) is
    the probability of accepting a step from $\vec{\theta_i}$ to
    $\vec{\theta_j}$.}.}
\item{Once we confirm that the auto-correlation time has converged
  (i.e., the MCMC algorithm has also converged), we continue sampling
  until the effective sample size becomes
  $\geq$100,000.\footnote{Sample number of $10^5$ corresponds to an
  uncertainty of only $\sim$3\%\ for the distribution of the
  99.73\%\ highest density interval (HDI).}}
\item{We finally obtain a probability distribution and its mode (i.e.,
  the best fit value) for each parameter by projecting the total
  probability density distribution on each parameter axis.}
\end{itemize}

\subsection{Comparison of MCMC and $\chi^2$ methods} \label{sec:comp}
\begin{figure}
  \begin{center}   
    \includegraphics[width=9cm,angle=0]{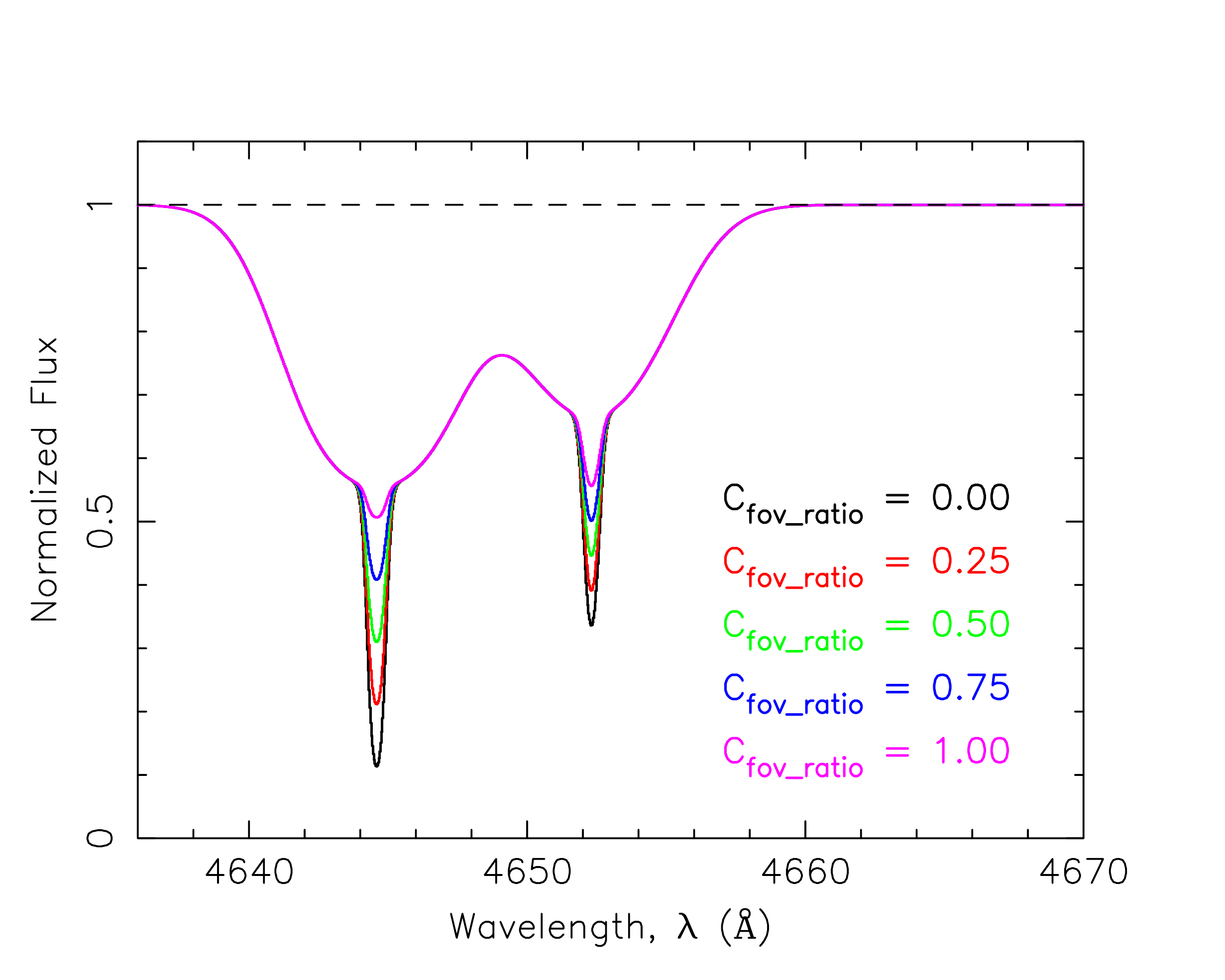}
  \end{center}
  \caption{Synthesized spectra with two \ion{C}{4} absorption
    components whose line parameters are log($N$/\cmm) = 15, $b$ =
    200~\kms, and \cf\ = 0.5 for the broader component, and
    log($N$/\cmm) = 14, $b$ = 20~\kms, and \cf\ = 0.5 for the narrower
    component.  The spectra with different \cfratio\ are overlaid in
    different colors. Even a synthesize spectrum with no overlap
    (i.e., black curve) does not reach zero flux at its center, since
    optical depth of two absorption components are not large enough to
    be saturated.\label{fig:cfovmodel}}
\end{figure}

We compare the fitting efficiencies of the MCMC method and the
traditional $\chi^2$ method as follows. For the latter, we use {\tt
  MINFIT}.  First, we synthesize spectra with two \ion{C}{4}
absorption components whose line widths and column densities are close
to the broad and narrow components in the mini-BAL system that we will
discuss in the next section: $b$ = 200~\kms\ and log($N$/\cmm) = 15
for the former and $b$ = 20~\kms\ and log($N$/\cmm) = 14 for the
latter, but both components have the same absorption redshift and
covering factor (\zabs\ = 2.0 and \cf\ = 0.5). These parameters are
kept fixed. But we do change overlap conditions as \cfnorm\ = 0, 0.25,
0.5, 0.75, and 1. A spectrum is synthesized from 4634\AA\ to
4661\AA\ with a pixel size of 0.03\AA\ pixel$^{-1}$. As shown in
Figure~\ref{fig:cfovmodel}, the total absorption depth is deeper when
\cfratio\ is smaller.  Next, we convolve the spectrum with the
appropriate line-spread function to simulate a spectral resolution of
$\lambda/\Delta\lambda$ = 36,000, which is a typical values for
observational data as is presented in Section~3.  In fact, the
convolution does not significantly affect the absorption profiles
since both the broad and the narrow components are much broader than
the spectral resolution element ($\Delta v$ $\sim$ 8.3~\kms).  To
examine how much the fitting accuracy would improve as the S/N
increases, we synthesize spectra with different noise realizations
with S/N from 10 to 100 pixel$^{-1}$ in steps of 10 and reproduce the
fit parameters repeatedly.

Once all the spectra are synthesized, we fit them with VP components
using {\tt mc2fit} and {\tt MINFIT}, respectively. We repeat this
analysis 10 times by changing the seed value of the random number
generator (used for adding noise). Finally, we calculate the average
and the standard deviation of best-fit (i.e., mode) values for each
line parameter.

\begin{figure*}
  \begin{center}
    \includegraphics[width=6.5cm,angle=0]{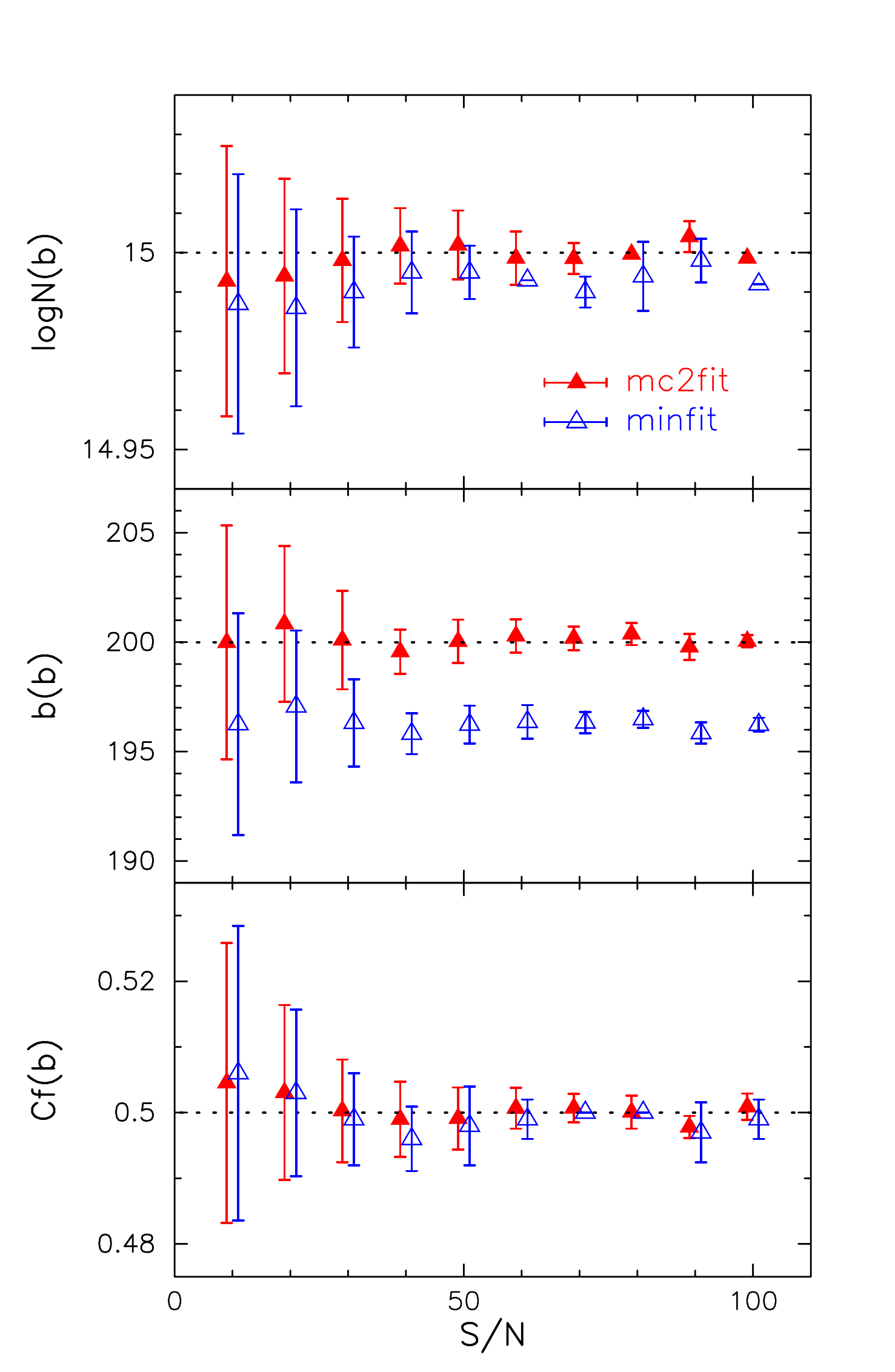}
    \includegraphics[width=6.5cm,angle=0]{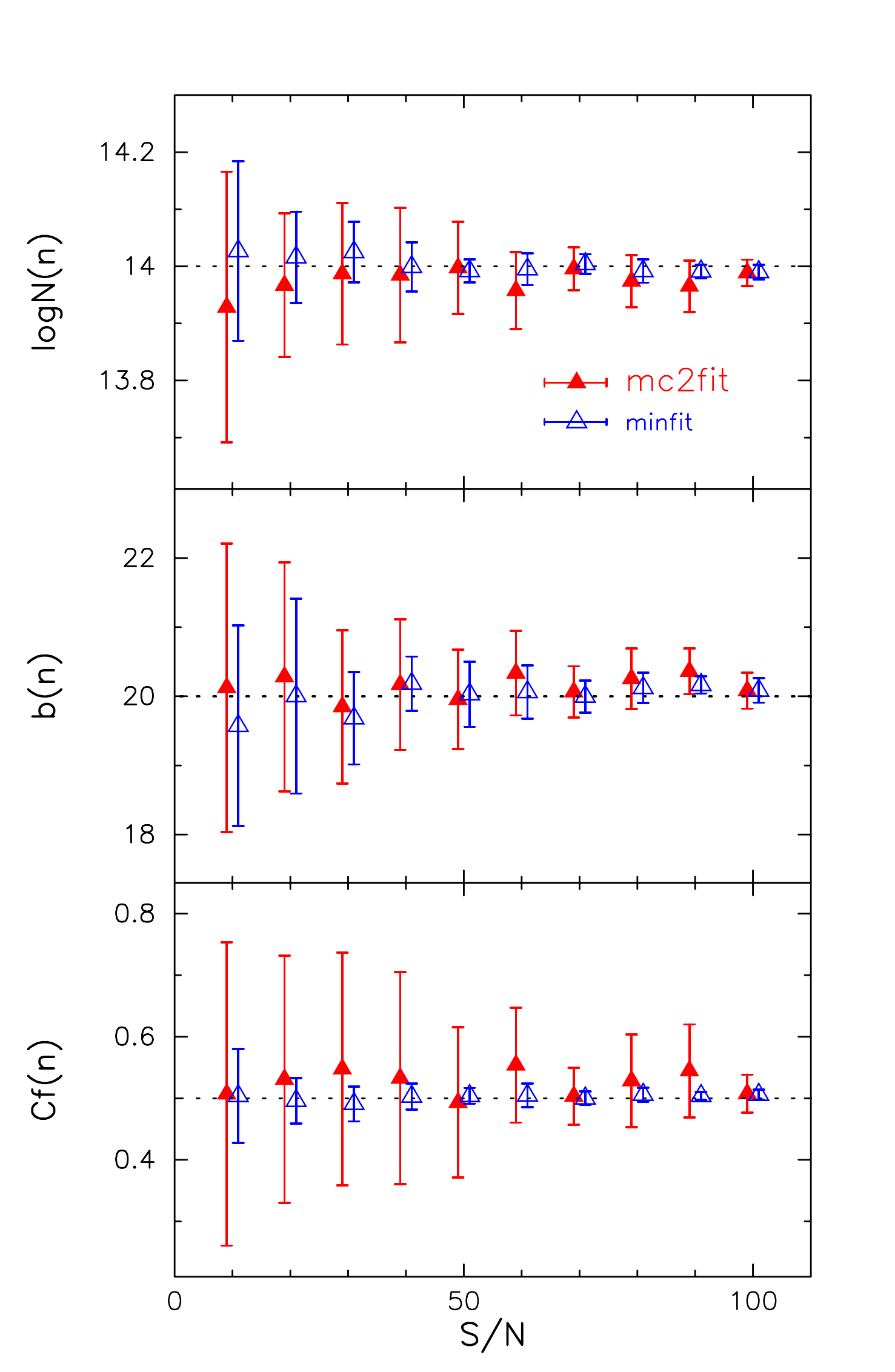}
 \end{center}
  \caption{Comparison of fit parameters for the synthesized spectra in
    Figure~\ref{fig:cfovmodel} (\logN, $b$, and \cf\ from top to
    bottom) using {\tt mc2fit} (red filled triangle) and {\tt MINFIT}
    (blue open triangle) with 1$\sigma$ uncertainties (vertical solid
    line) as a function of S/N. The normalized covering factor is
    fixed to \cfnorm\ = 0.5. To facilitate comparison, the points are
    intentionally shifted horizontally. Horizontal dotted lines denote
    correct values. Left and right figures are results for the broad
    (b) and the narrow (n) components,
    respectively.\label{fig:SNsim1}}
\end{figure*}

Figure~\ref{fig:SNsim1} shows a comparison of fit parameters (\logN,
$b$, and \cf) that are returned by {\tt mc2fit} and {\tt MINFIT} as a
function of S/N. In this demonstration, we adopt \cfnorm\ = 0.5 (i.e.,
\cfov\ = 0.25 and \cfratio\ = 0.5)\footnote{Here, min(\cfov) = 0 and
  max(\cfov) = 0.5 since \cf$_{(1)}$ = \cf$_{(2)}$ = 0.5. By
  substituting these into equations~\ref{eqn:cfnorm} and
  \ref{eqn:cfratio}, we obtain \cfnorm = $C_{\rm fov\_ratio}$ =
  $C_{\rm fov}/0.5$. Therefore, $C_{\rm fov}$ = 0.25 and $C_{\rm
    fov\_ratio}$ = 0.5 when \cfnorm = 0.5.}.  The code {\tt mc2fit}
always provides the correct values within 1$\sigma$ uncertainty, while
{\tt MINFIT} underestimates \logN\ and $b$ for the broad component
even in high-quality spectrum with S/N = 100 pixel$^{-1}$. The
discrepancy is probably a result of not allowing for an overlap
covering factor in {\tt MINFIT}. Thus, for cases where multiple
absorbers overlap along the line of sight, the MCMC method is required
to derive correct line parameters.\footnote{Here, we emphasize that
  \cf\ is reliable even if we use {\tt MINFIT}. Therefore, the past
  results of classifying absorption lines into intrinsic and
  intervening ones base on values of \cf\ determined by $\chi^2$
  methods are still reliable.}

\begin{figure*}
  \begin{center}
    \includegraphics[width=6.5cm,angle=0]{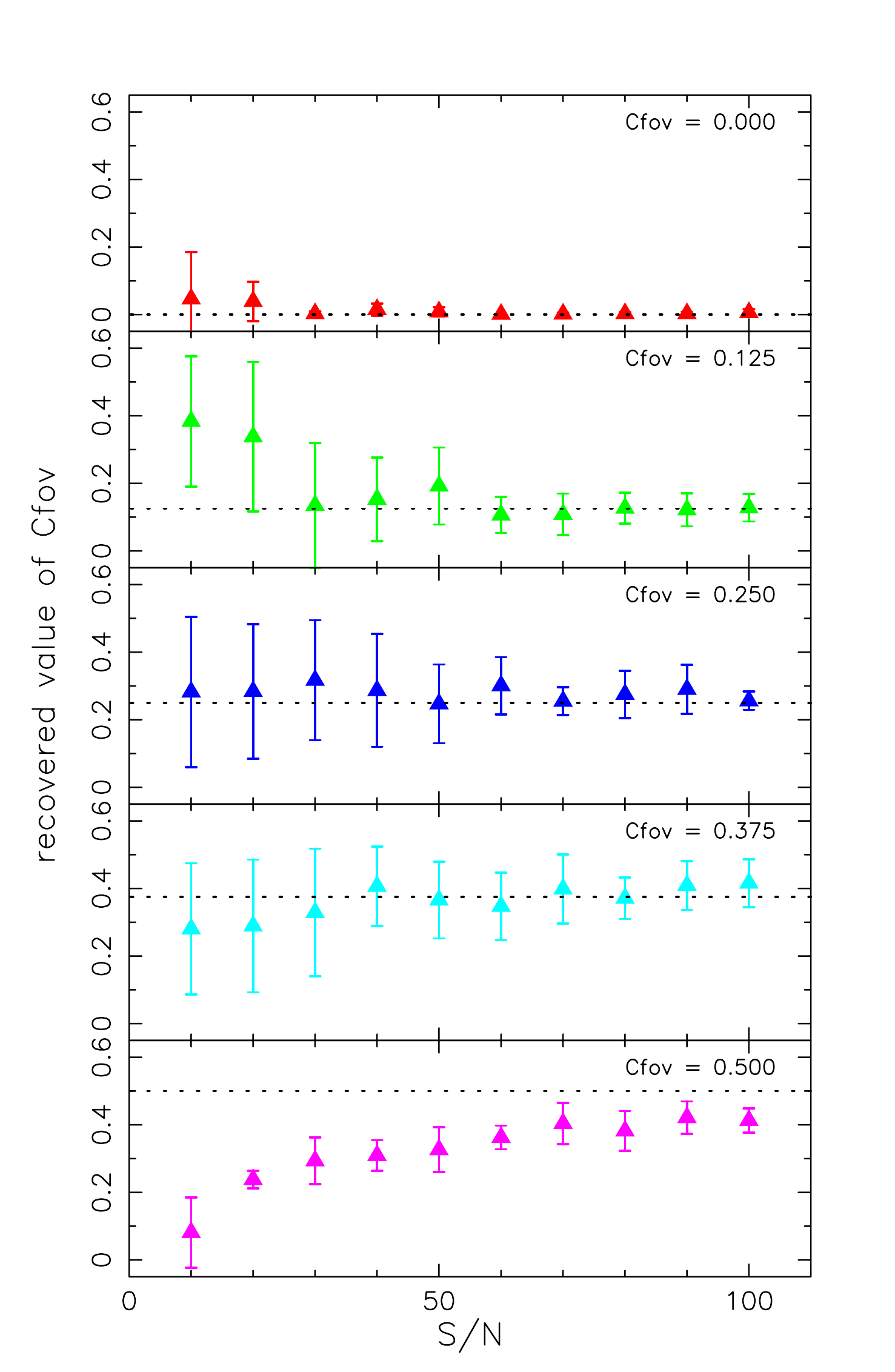}
    \includegraphics[width=6.5cm,angle=0]{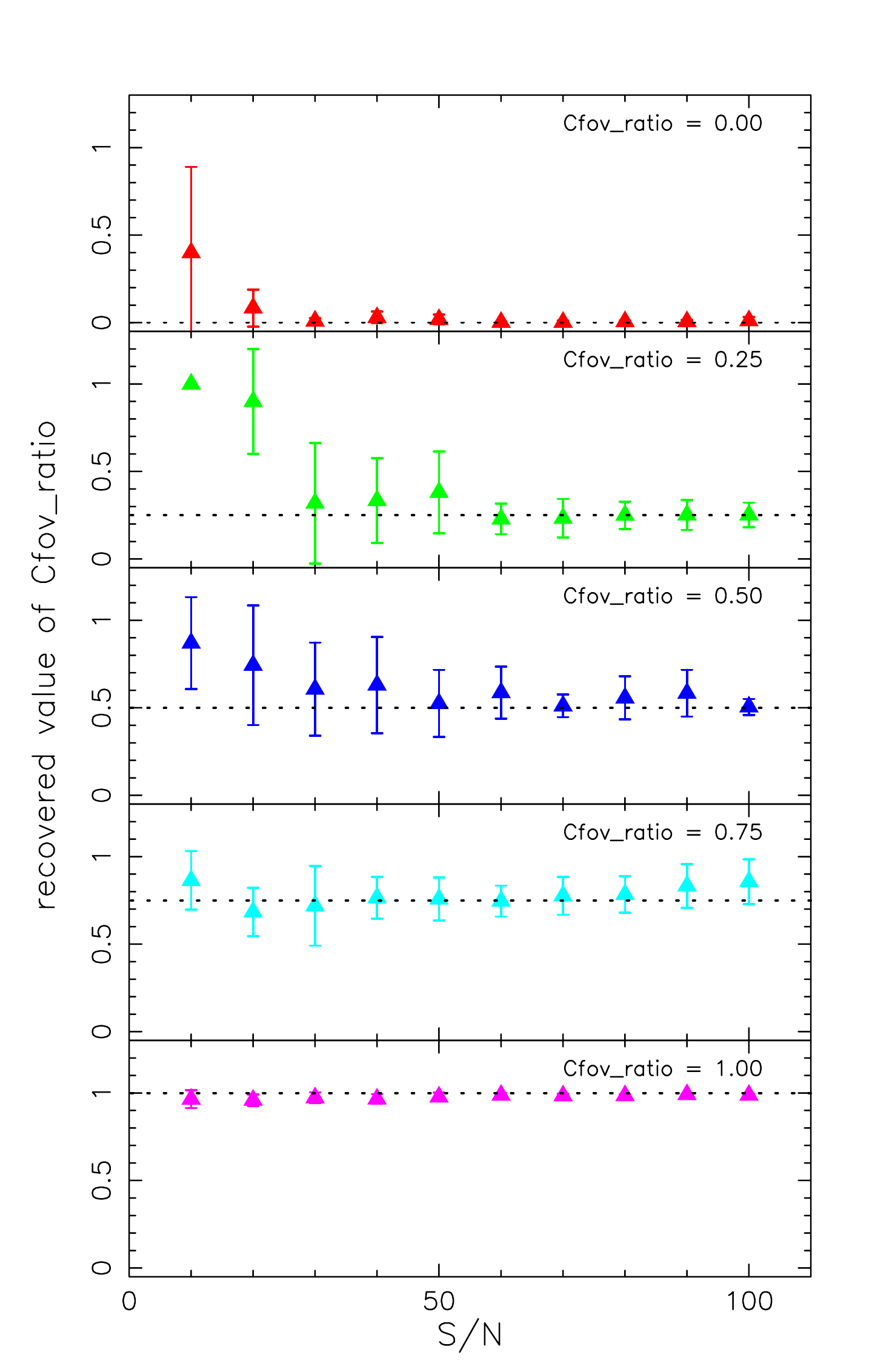}
 \end{center}
  \caption{Overlap covering factor (\cfov, left) and overlap covering
    factor ratio (\cfratio, right) that are reproduced by {\tt mc2fit}
    (filled triangle) with 1$\sigma$ uncertainties (vertical solid
    line) as a function of S/N.  The normalized covering factor is
    fixed as \cfnorm\ = 0.0, 0.25, 0.5, 0.75, and 1.0 from top to
    bottom (i.e., \cfov\ = 0.0, 0.125, 0.25, 0.375, 0.5 and
    \cfratio\ = 0.0, 0.25, 0.5, 0.75, 1.0).  Horizontal dotted lines
    denote correct values.\label{fig:SNsim2}}
\end{figure*}

\begin{figure}
  \begin{center}   
    \includegraphics[width=8.5cm,angle=0]{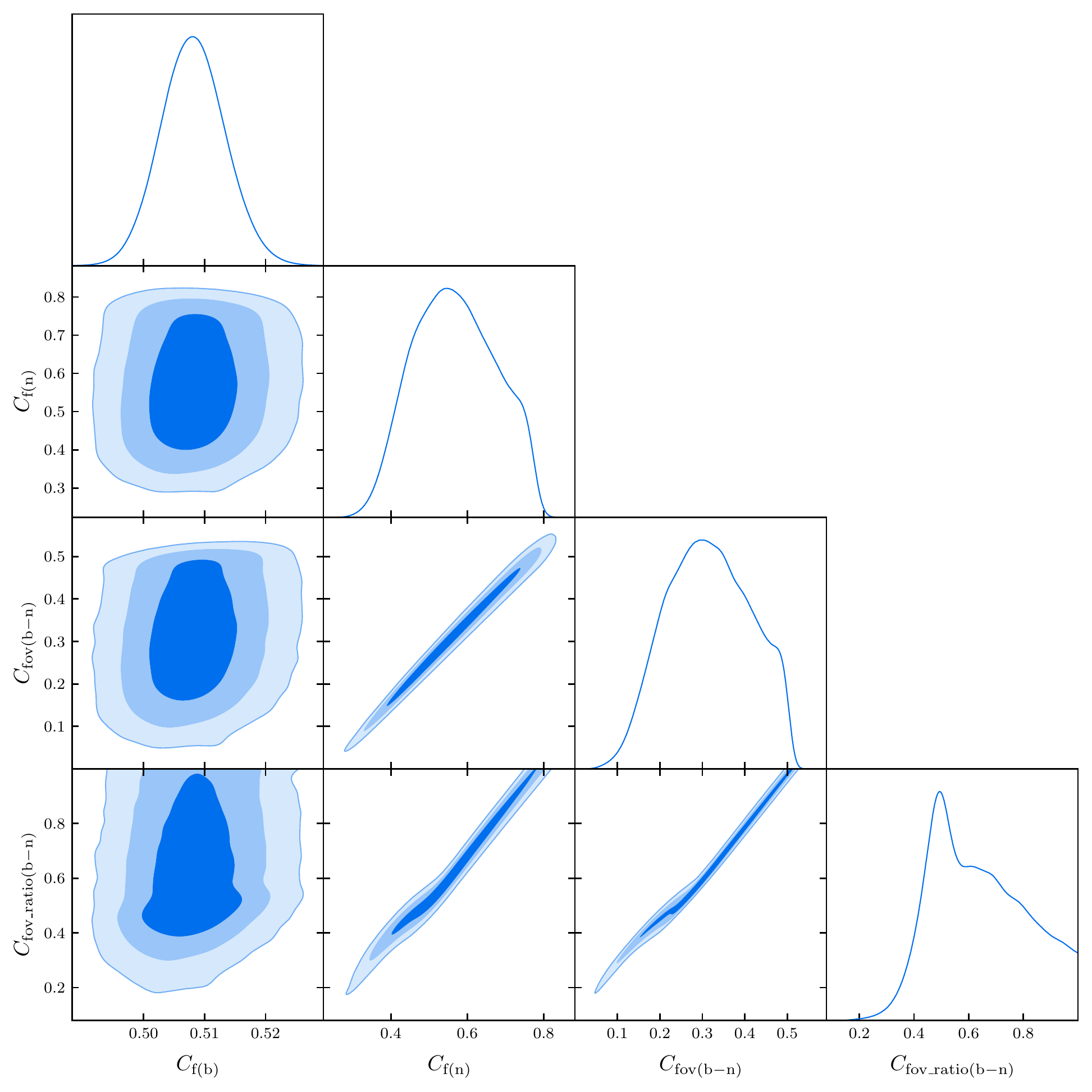}
  \end{center}
  \caption{Example of posterior probability distribution function of
    covering factors (\cf, correct values are 0.5) of both broad and
    narrow components as well as their overlap covering factor (\cfov,
    0.25) and overlap covering factor ratio (\cfratio,
    0.5).\label{fig:corner}}
\end{figure}

We also summarize the fit results of the overlap covering factor
(\cfov) and the overlap covering factor ratio (\cfratio) by {\tt
  mc2fit} in Figure~\ref{fig:SNsim2} for some selected values,
\cfov\ = 0.0, 0.125, 0.25, 0.375, 0.5, and \cfratio\ = 0.0, 0.25, 0.5,
0.75, 1.0.  An example of a corner plot is also shown in
Figure~\ref{fig:corner}. The code recovers \cfratio\ generally well,
while \cfov\ is always under-estimated even in spectra with S/N $\sim$
100~pixel$^{-1}$ if the correct \cfov\ value is 0.5.  Therefore, we
will use \cfratio\ to examine whether absorbers overlap, since it is
always determined reliably (at least the correct value is within
1$\sigma$ uncertainty) if the data quality is high enough, S/N $\geq$
30~pixel$^{-1}$.

\section{Application to Observed Spectra} \label{sec:um675}
To test {\tt mc2fit} further, we apply it to a mini-BAL system whose
line profile is easy to separate into multiple absorption components
(i.e., an ideal target to test the code) that is detected in an
optically bright quasar. There also exist several narrow absorption
line systems in the same quasar spectrum.

\subsection{Mini-BAL Quasar UM675}
In the spectrum of the radio-loud\footnote{We classify UM675 as a
  radio-loud quasar with a radio-loudness parameter of $\mathcal{R}$ =
  351, using an optical magnitude ($V$ = 17.4) and a radio flux
  ($f_\nu$ = 122~mJy at 4.85~GHz) from \citet{gri94}, while it was
  classified as a radio-quiet quasar in \citet{ham95,ham97b}.}
luminous quasar UM675 at \zem\ = 2.147, \citet{sar88} detected three
\ion{C}{4} NALs at \zabs\ $\sim$ 1.7666, 1.9288, and 2.0083.
\citet{ham95} also discovered \ion{C}{4}, \ion{N}{5}, and
\lya\ mini-BALs with FWHM $\sim~500~\kms$ with a high metallicity ($Z$
$>$ 2$Z_{\odot}$) at \zabs\ $\sim$ 2.134 corresponding to an offset
velocity of \voff\ $\sim$1500~\kms\ from the quasar emission
redshift. The mini-BAL system is optically thin because the
Lyman-continuum edge is absent.  The system also has a wide variety of
absorption lines with a full range of ionization from \ion{Ne}{8}
(ionization potential IP = 239~eV) to \ion{C}{3} (IP = 48~eV)
\citep{bea91,ham95}, which suggests that the gas density and/or radial
distance from the flux source span a factor of $\sim$100 or $\sim$10,
respectively.

Using intermediate-resolution spectra ($R$ $\sim$ 1500), \citet{ham95}
discovered time variability in the absorption strength of the
\ion{C}{4} and \ion{N}{5} mini-BALs in $\Delta t_{\rm rest} \sim
2.9$~years in the quasar rest-frame, and placed a lower limit on the
electron density of $n_{\rm e} \geq 4000$~cm$^{-3}$ and an upper limit
on the radial distance from the source of $r \leq 200$~pc, assuming
the absorber is in ionization equilibrium.  There is clear evidence
for partial coverage of the background flux source (\cf\ =
0.63$^{+0.2}_{-0.1}$ for \ion{N}{5} and 0.34$^{+0.1}_{-0.0}$ for
\ion{C}{4} at \voff\ $\sim$1500~\kms; \citealt{ham97b}) based on high
resolution spectra taken with Keck/HIRES, which means the mini-BAL
system is physically associated with the quasar (i.e., {\it intrinsic}
QAL). \citet{ham97b} also detected three narrow components with FWHM
$\sim$ 30--50~\kms\ inside the smooth mini-BAL profiles of \ion{C}{4}
and \ion{N}{5}, as shown in Figure~2 of their paper.

\begin{figure*}
 \begin{center}
  \includegraphics[width=17cm,angle=0]{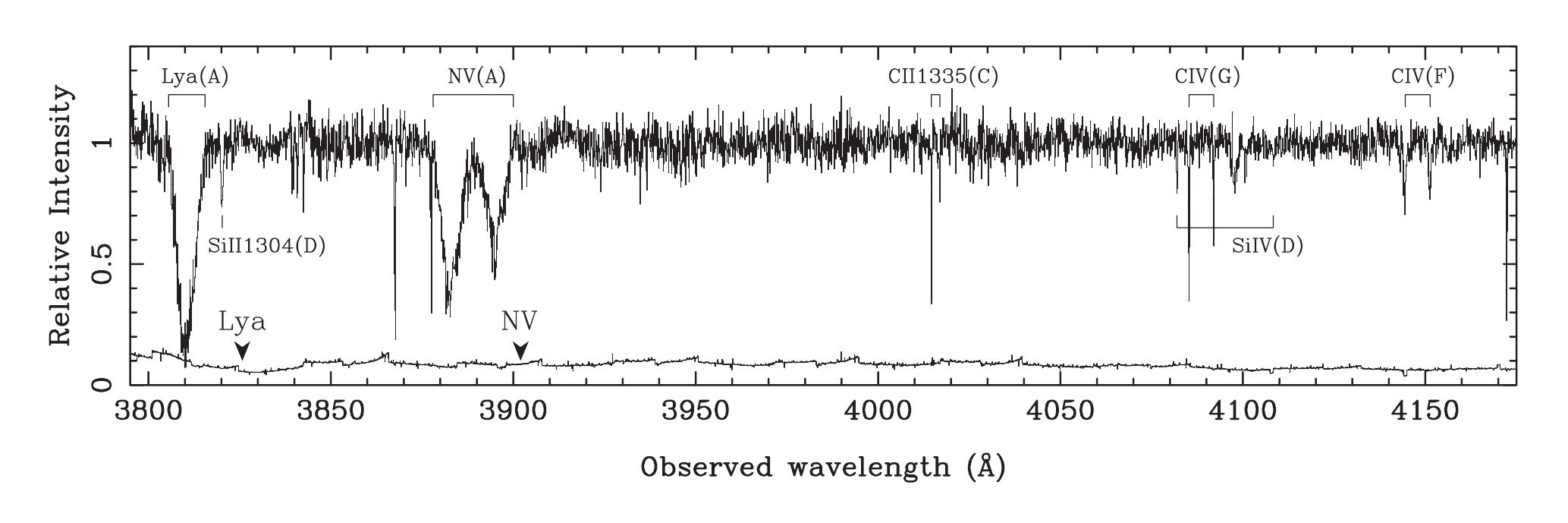}
  \includegraphics[width=17cm,angle=0]{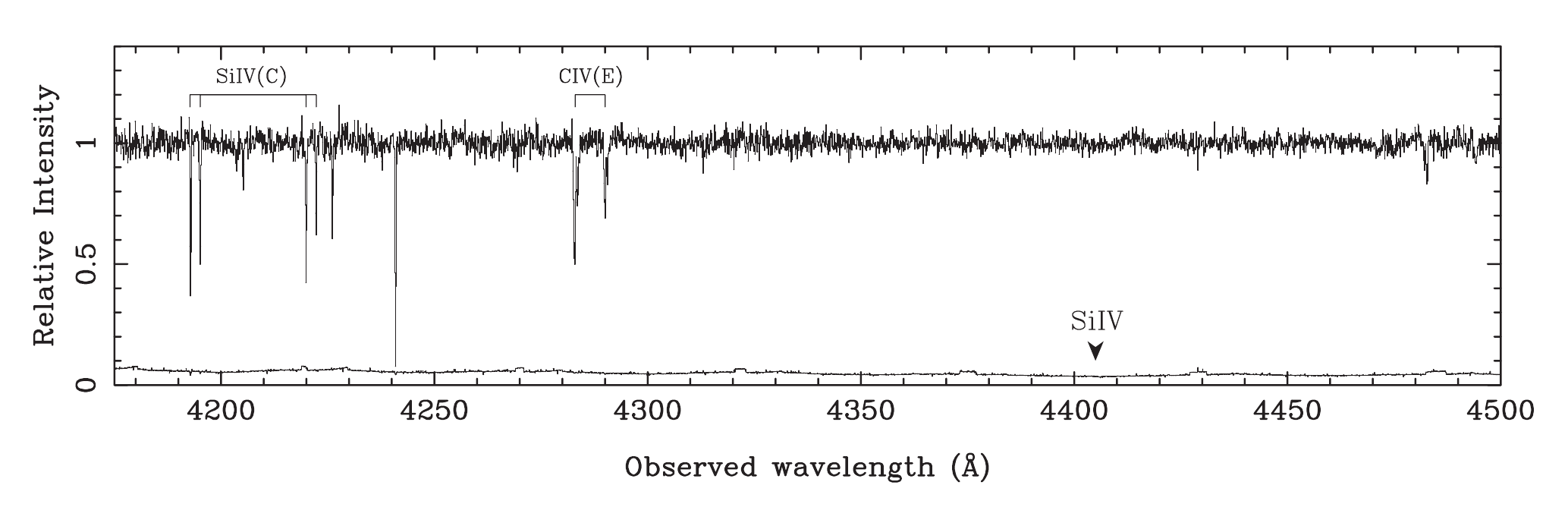}
  \includegraphics[width=17cm,angle=0]{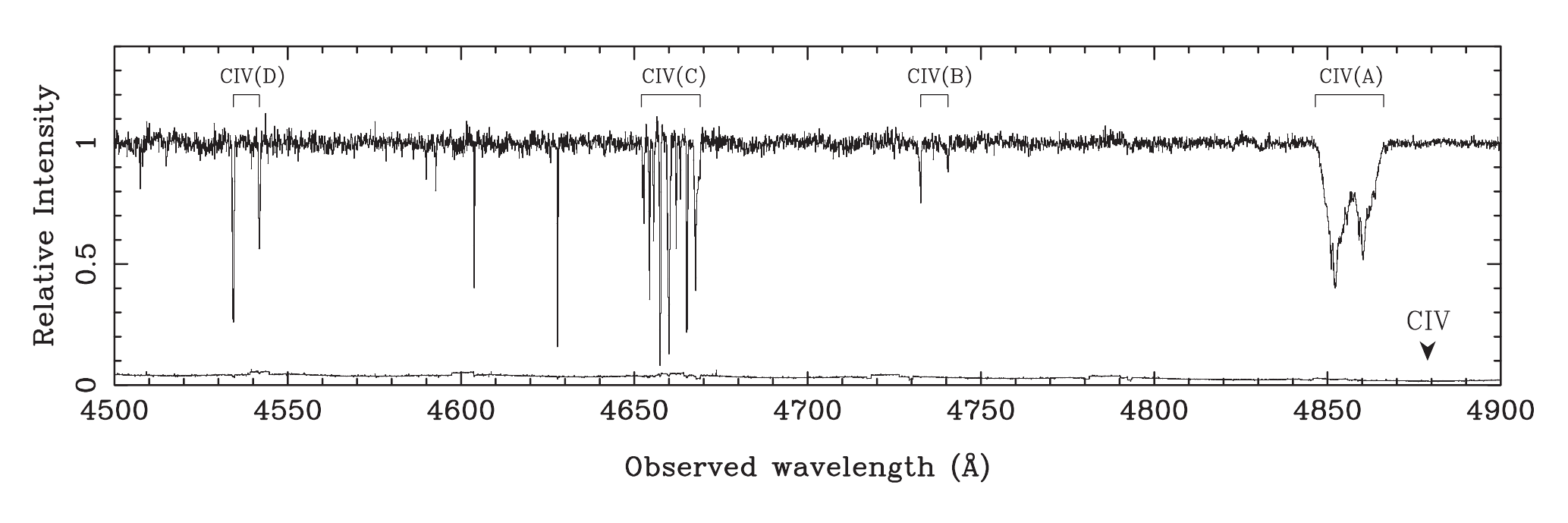}
 \end{center}
\caption{Normalized spectrum of UM675 after rebinning to 0.1~\AA\ per
  pixel. The \ion{C}{4}, \ion{N}{5}, \ion{Si}{4} doublets and other
  detected single lines are marked. Positions of the quasar emission
  lines of \lya, \ion{N}{5}, \ion{Si}{4}, and \ion{C}{4} are marked
  with downward arrows. The regions blueward of the \lya\ absorption
  in system~A and redward of the \ion{C}{4} emission line are not
  shown. The lower trace shows the 1$\sigma$ error
  spectrum.\label{fig:UM675}}
\end{figure*}

To date, high-resolution spectra ($R$ $\sim$ 35000 -- 45000) of the
quasar have been obtained three times with Keck/HIRES in 1994
September (hereafter, epoch~E1), 1998 September (E2), and 2008 January
(E4), and once with Subaru/HDS in 2005 August (E3) as summarized in
Table~\ref{tab:obslog}.  The spectrum in epoch~E1 is shown here in
Figure~\ref{fig:UM675} as an example.  Using the high-resolution
spectra in epochs~E1 and E3, \citet{mis14} confirmed again that the
absorption strengths (i.e., equivalent width) of \lya, \ion{C}{4}, and
\ion{N}{5} show an obvious time variability with $>$5$\sigma$
significance level, but with no discernible changes in the narrow
components inside it.  In this study, we use all of the above that are
available from the Keck Observatory Archive (KOA)\footnote{\tt
  http://www2.keck.hawaii.edu/koa/public/koa.php} or obtained
by us with Subaru/HDS although the \ion{N}{5} mini-BAL is not
covered by the spectrum in epoch E2. We reduced the data ourselves
using a data reduction pipeline for Keck/HIRES data (MAuna Kea Echelle
Extraction; MAKEE)\footnote{\tt
  http://www.astro.caltech.edu/\textasciitilde tb/makee/index.html}.
After normalizing the spectra, we searched for absorption lines whose
depth is greater than 5 times the corresponding noise level. We
identified six \ion{C}{4} NALs at \zabs\ = 2.0569, 2.0083, 1.9288,
1.7663, 1.6769, and 1.6387 between the \lya\ and \ion{C}{4} emission
lines (Systems~B, C, D, E, F and G, hereafter) in addition to the
mini-BAL system at \zabs\ $\sim$ 2.134 (System~A, hereafter), as
summarized in Table~\ref{tab:abs}. Among these, System~C, D, and E
were already reported in \citet{sar88}.

\begin{deluxetable*}{ccccccccc}
\tablecaption{Log of Monitoring Observations\label{tab:obslog}}
\tablewidth{0pt}
\tablehead{
\colhead{Epoch}                       &
\colhead{Obs. Date}                   &
\colhead{Instrument}                  &
\colhead{$\lambda$-coverage}          &
\colhead{Exp. Time}                   &
\colhead{$\lambda / \Delta\lambda^a$} &
\colhead{S/N (\ion{N}{5})$^b$}        &
\colhead{S/N (\ion{C}{4})$^b$}        &
\colhead{Source of Data$^c$}          \\
\colhead{}                   &
\colhead{}                   &
\colhead{}                   &
\colhead{(\AA)}              &
\colhead{(sec.)}             &
\colhead{}                   &
\colhead{(pixel$^{-1}$)}      &
\colhead{(pixel$^{-1}$)}      &
\colhead{}                   \\
}
\startdata
E1 & 1994 Sep 24--25 & Keck+HIRES & 3750--6100 & 18000 & 34,000 & 47   &  9 & A \\
E2 & 1998 Sep 22     & Keck+HIRES & 4150--6520 &  3600 & 47,800 & $^d$ & 16 & B \\
E3 & 2005 Aug 19--20 & Subaru+HDS & 3600--5980 & 23000 & 36,000 & 48   & 28 & C \\
E4 & 2008 Jan 15     & Keck+HIRES & 3200--5990 &   840 & 47,750 & 12   &  9 & B 
\enddata
\tablenotetext{a}{Spectral resolution.}
\tablenotetext{b}{S/N on the red side of \ion{N}{5}
  ($\sim$3950\AA) and \ion{C}{4} ($\sim$4900\AA) mini-BALs.}
\tablenotetext{c}{(A) Provided by Fred Hamann, (B) from Keck
  Observatory Archive (KOA), and (C) taken by us (proposal ID is
  S05A-041).}
\tablenotetext{d}{\ion{N}{5} is not covered by the observed spectrum.}
\end{deluxetable*}

\begin{deluxetable*}{cccccc}
\tablecaption{Absorption Line Systems\label{tab:abs}}
\tablewidth{0pt}
\tablehead{
\colhead{System}          &
\colhead{\zabs}           &
\colhead{\voff$^a$}       &
\colhead{Class$^b$}       &
\colhead{Ions$^c$}        &
\colhead{Variability$^d$} \\
\colhead{}                &
\colhead{}                &
\colhead{(\kms)}          &
\colhead{}                &
\colhead{}                &
\colhead{}                \\
}
\startdata
 A & $\sim$2.134  & $\sim$1,240 & mini-BAL & \ion{C}{4}, \ion{N}{5}, (\lya)         & Y \\
 B &       2.0569 &       8,710 &      NAL & \ion{C}{4}                             & N \\
 C$^e$ &   2.0083$^f$ &   13,510 &      NAL & \ion{C}{4}, \ion{Si}{4}, (\ion{C}{2})  & N \\
 D &       1.9288 &      21,520 &      NAL & \ion{C}{4}, \ion{Si}{4}, (\ion{Si}{2}) & N \\
 E &       1.7663 &      38,470 &      NAL & \ion{C}{4}                             & N \\
 F &       1.6769 &      48,120 &      NAL & \ion{C}{4}                             & N \\
 G &       1.6387 &      52,310 &      NAL & \ion{C}{4}                             & N 
\enddata
\tablenotetext{a}{Offset velocity from the quasar emission redshift
  (\zem = 2.147). Positive values denote blueshifted from the quasar.}
\tablenotetext{b}{Absorption class: mini-BAL or NAL.}
\tablenotetext{c}{Detected ions in each absorption system. Ions in
  parentheses represent transitions that are not fitted because they are
  not doublets.}
\tablenotetext{d}{Absorption line is variable (Y) or not (N).}
\tablenotetext{e}{Multiple components are separated from each other by 
  unabsorbed spectral regions, but we regard them as a single system since
  their velocity separation is $<$200~\kms, following \citet{mis07a}.}
\tablenotetext{f}{Redshift of the middle of five components.}
\end{deluxetable*}

\subsection{MCMC fit to a mini-BAL and six NALs}
\begin{figure}
  \begin{center}   
    \includegraphics[width=9cm,angle=0]{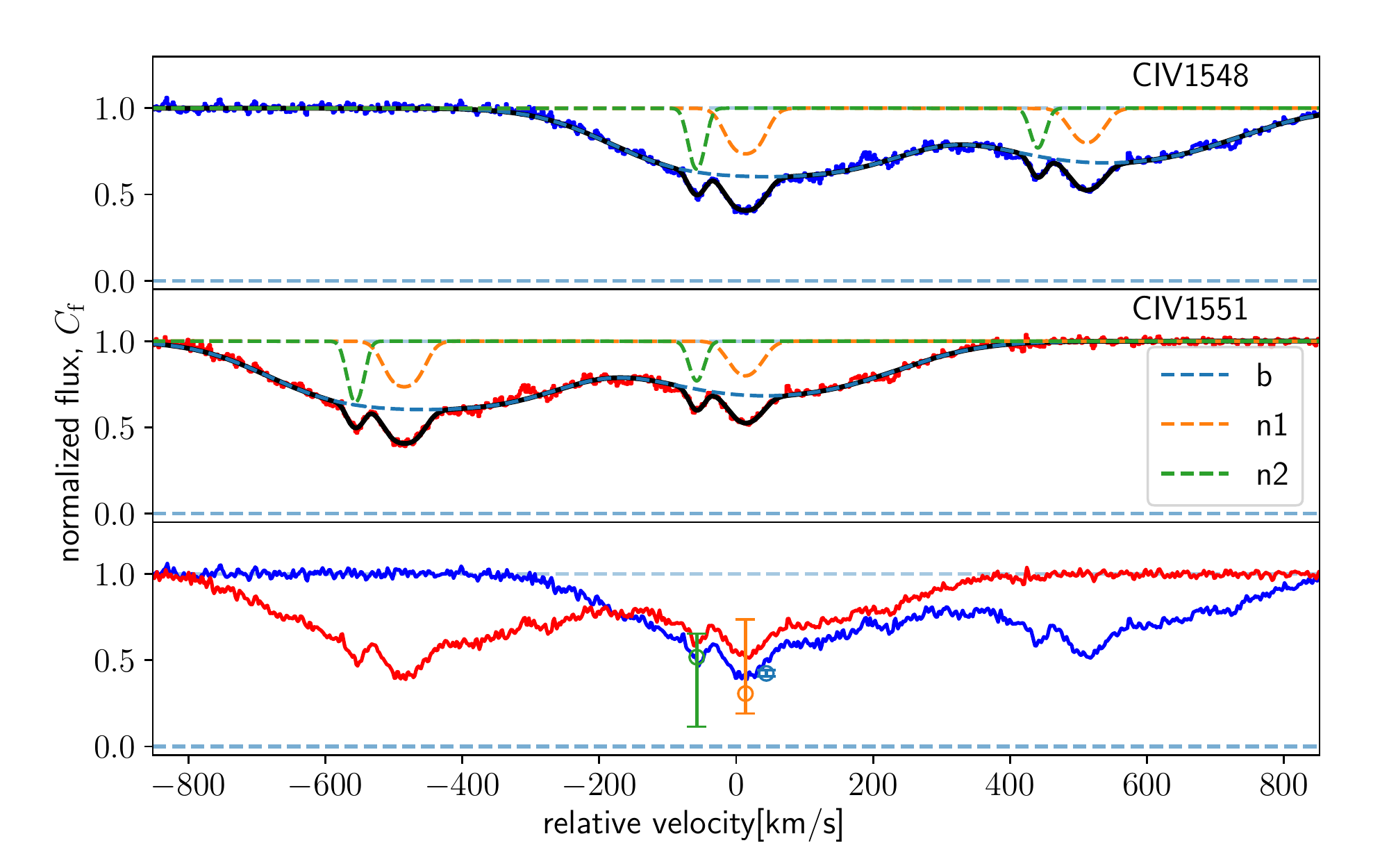}
    \includegraphics[width=9cm,angle=0]{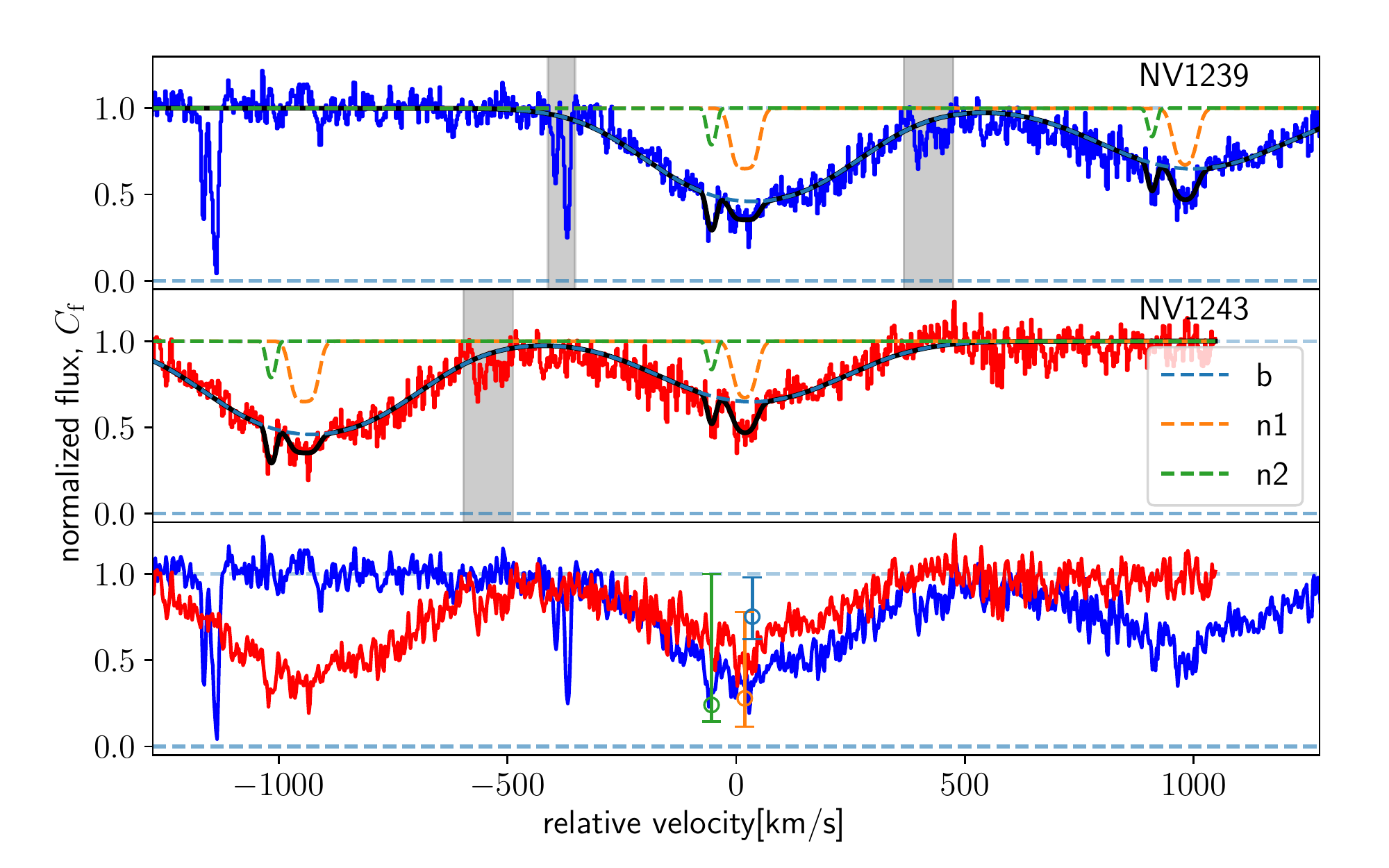}
  \end{center}
  \caption{Best-fit models for the \ion{C}{4} and \ion{N}{5} mini-BALs
    in System~A using the MCMC method. The velocity on the horizontal
    axis is defined relative to the flux-weighted center of the
    absorption profiles.  The top and middle panels show the profiles
    of the blue and red members of a doublet on a common velocity
    scale, with the model profile produced by {\tt mc2fit} superposed:
    dashed lines for each component and a solid line for the
    contribution of all components.  The bottom panel shows the two
    profiles together, along with the resulting covering factors (open
    circles) with their 1$\sigma$ error bars. Profiles in the shaded
    areas are ignored for fitting. The complete figure set is
    available in Appendix.\label{fig:bestfit}}
\end{figure}

We use {\tt mc2fit} to fit the \ion{C}{4} and \ion{N}{5} mini-BALs and
six \ion{C}{4} NALs in the spectra of UM675 in the four epochs.  As a
prior, we assume a uniform distribution between the lower and upper
limits of each parameter: [1.469, 2.147] (i.e., between the redshift
of \ion{C}{4} absorption line corresponding to the \lya\ emission line
and the quasar emission redshift) for \zabs, (0, 16] for
  log($N$/\cmm), (0, 2000) (i.e., the boundary between BAL and
  mini-BAL) for $b$~[\kms], (0, 1] for \cf, and (0, 1] for \cfnorm,
      where parentheses, (a, b), and square brackets [a, b] denote an
      interval from a to b, but only the latter includes the values of
      a and b themselves. The best-fit models are presented in
      Figure~\ref{fig:bestfit}.

We confirm that only the mini-BAL system (i.e., System~A) shows
statistically significant partial coverage in both \ion{C}{4} and
\ion{N}{5} profiles, while none of the NAL systems show partial
coverage with $> 3\sigma$ confidence level (i.e., the NAL systems are
probably intervening QALs)\footnote{We also confirm that any NAL
systems do not show variability in their equivalent width with $>$
3$\sigma$ level during the monitoring period, which also suggests
these are intervening NALs.}.  Therefore, we will study the fit
parameters\footnote{In this study, we do not discuss variability of
\zabs\ because it could be affected by small systematic
linear/nonlinear offsets of the spectra in each epoch from an
uncertainty in wavelength calibration.} only for System~A.  The best
fit parameters for System~A and their 1$\sigma$/3$\sigma$
uncertainties are summarized for all epochs in Table~\ref{tab:fit}.
Included are the absorption redshift, column density, Doppler
parameter, covering factor, and rest-frame equivalent width.  We
present the fitting results of the NAL systems (i.e., Systems~B -- G)
in Table~\ref{tab:fit2}, but only for epoch~E1 since there is no
variability.  In the following, we describe the results for 
each system in detail.

\subsection{System~A} \label{subsec:sysA}
\begin{figure}
  \begin{center}   
    \includegraphics[width=9cm,angle=0]{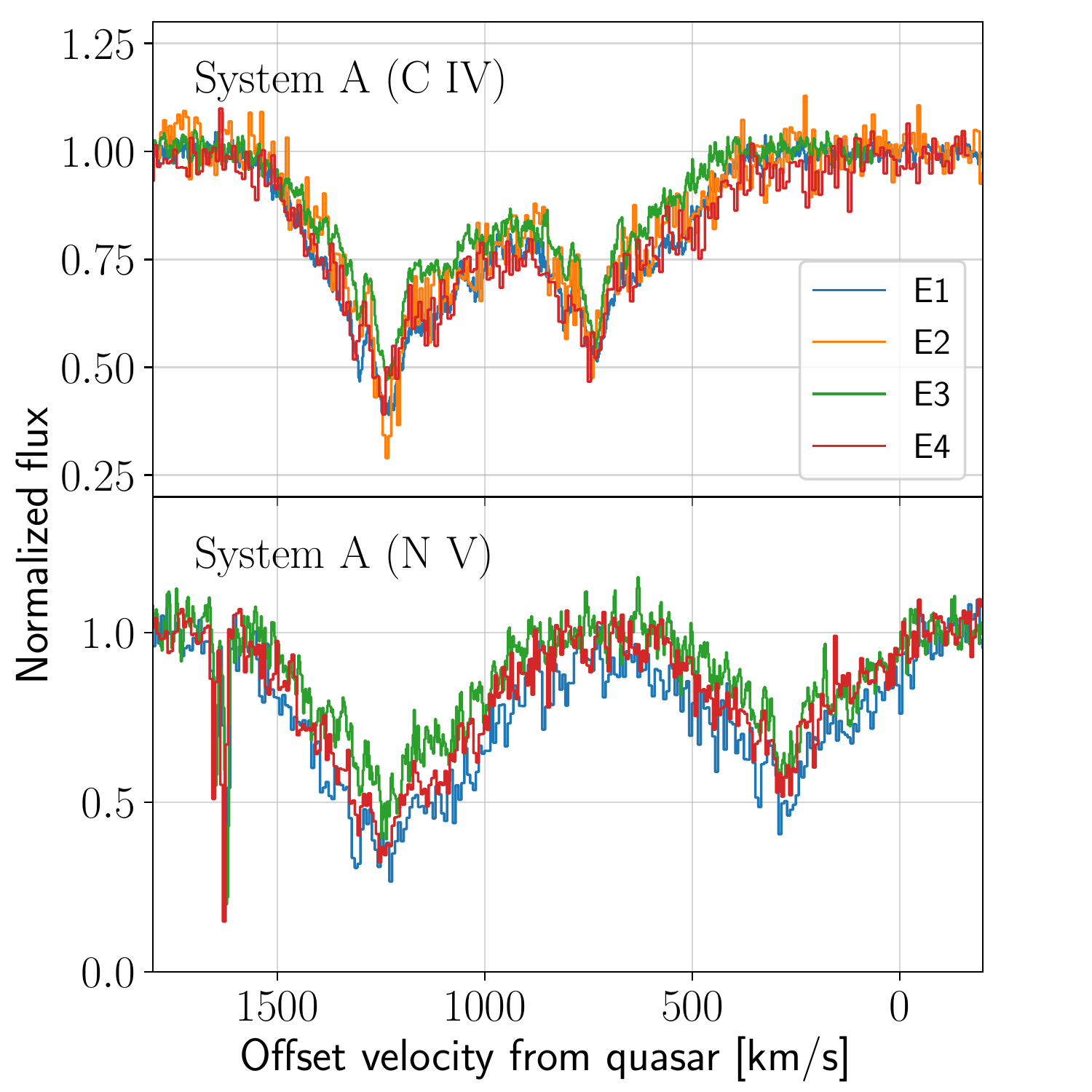}
  \end{center}
  \caption{Normalized spectra around the \ion{C}{4} (top) and
    \ion{N}{5} (bottom) mini-BALs in system~A. Vertical axis denotes
    normalized flux.  Horizontal axis denotes the offset velocity of the
    flux-weighted center of the {\it blue} member of the doublet
    (i.e., \ion{C}{4}~$\lambda$1548 and \ion{N}{5}~$\lambda$1239) from
    the quasar. The first to fourth epoch spectra are shown with blue,
    orange, green, and red histograms.\label{fig:sysA}}
\end{figure}

\begin{figure}
  \begin{center}   
    \includegraphics[width=8.5cm,angle=0]{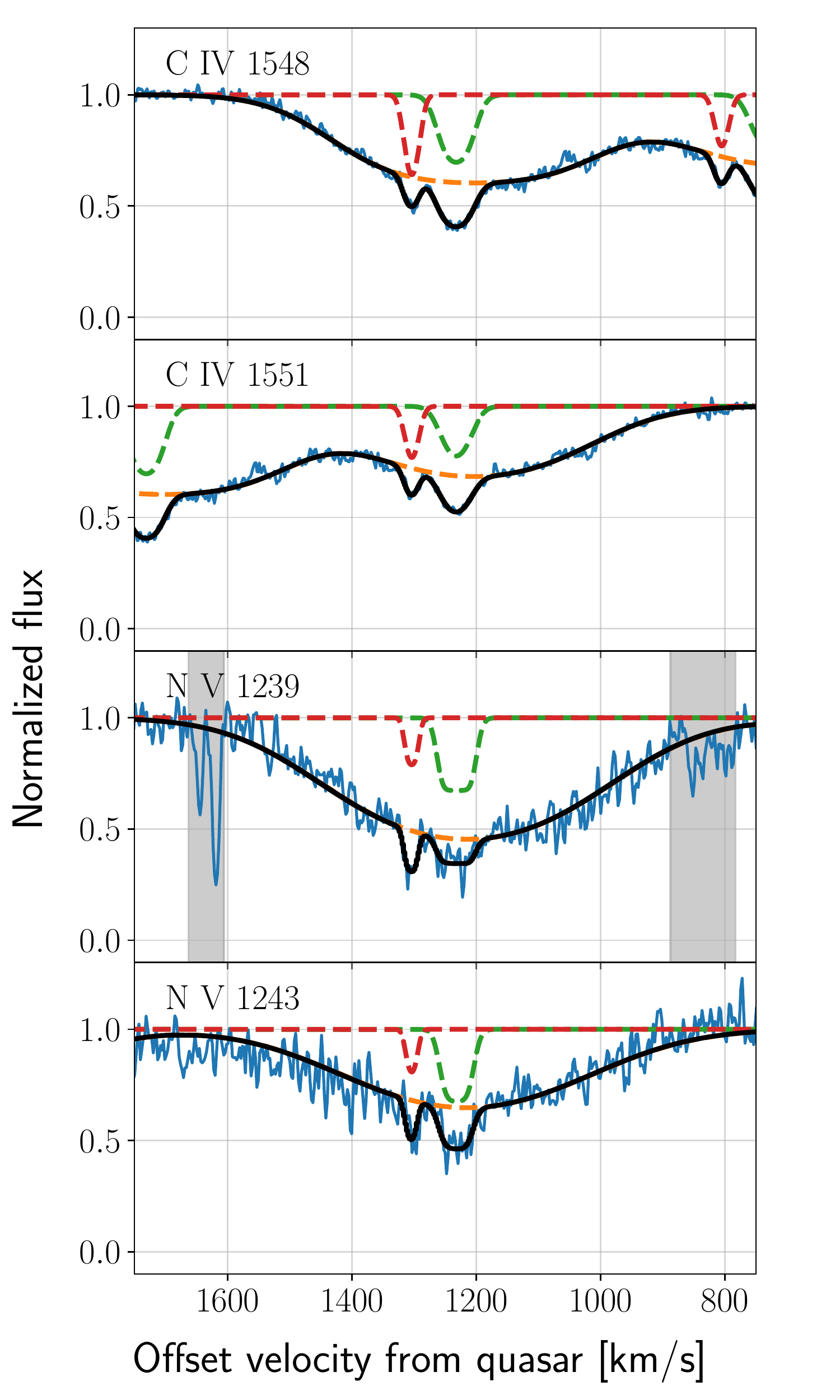}
  \end{center}
  \caption{Best fit model to the normalized \ion{C}{4} and \ion{N}{5}
    mini-BALs (System~A) in the quasar UM675 in epoch~E1. The observed
    (blue histogram) and modeled (black curve) spectra around
    \ion{C}{4}~1548, \ion{C}{4}~1551, \ion{N}{5}~1239, and
    \ion{N}{5}~1243 are shown from the top to the bottom as a function
    of offset velocity from the quasar emission redshift. Orange,
    green, and red dashed curves are the best models for components~b,
    n1, and n2, respectively.  The horizontal axis is offset velocity
    from the quasar emission redshift, while the vertical axis is
    normalized flux. Absorption lines in the shaded areas at
    \voff\ $\sim$ 850~\kms\ and 1650~\kms\ in the third panel are
    unrelated profiles.\label{fig:bestA}}
\end{figure}

We clearly detect \ion{C}{4} and \ion{N}{5} mini-BALs at \zabs\ $\sim$
2.134 but \ion{Si}{4} is not detected in the corresponding region
($\lambda$ $\sim$ 4400~\AA).  Figure~\ref{fig:sysA} shows normalized
spectra around the \ion{C}{4} and \ion{N}{5} mini-BALs for all four
epochs, which clearly indicate the system is variable.  Although
\citet{ham97b} detected three narrow components within a broad
absorption feature in the \ion{C}{4} and \ion{N}{5} mini-BALs in epoch
E1, we consider only two narrow components because the third (weakest)
component in \citet{ham97b} is very weak and almost swamped in the
noise around the \ion{N}{5} mini-BAL in the same epoch and in the
\ion{C}{4} mini-BAL in all other epochs.  Hereafter, we refer to them
as components~b (broad), n1 (narrow~1; at longer wavelength), and n2
(narrow~2; at shorter wavelength), respectively, as shown in
Figure~\ref{fig:bestA}, where the best fit model is superposed for
comparison.

\begin{figure*}
  \begin{center}
    \includegraphics[width=8cm,angle=0]{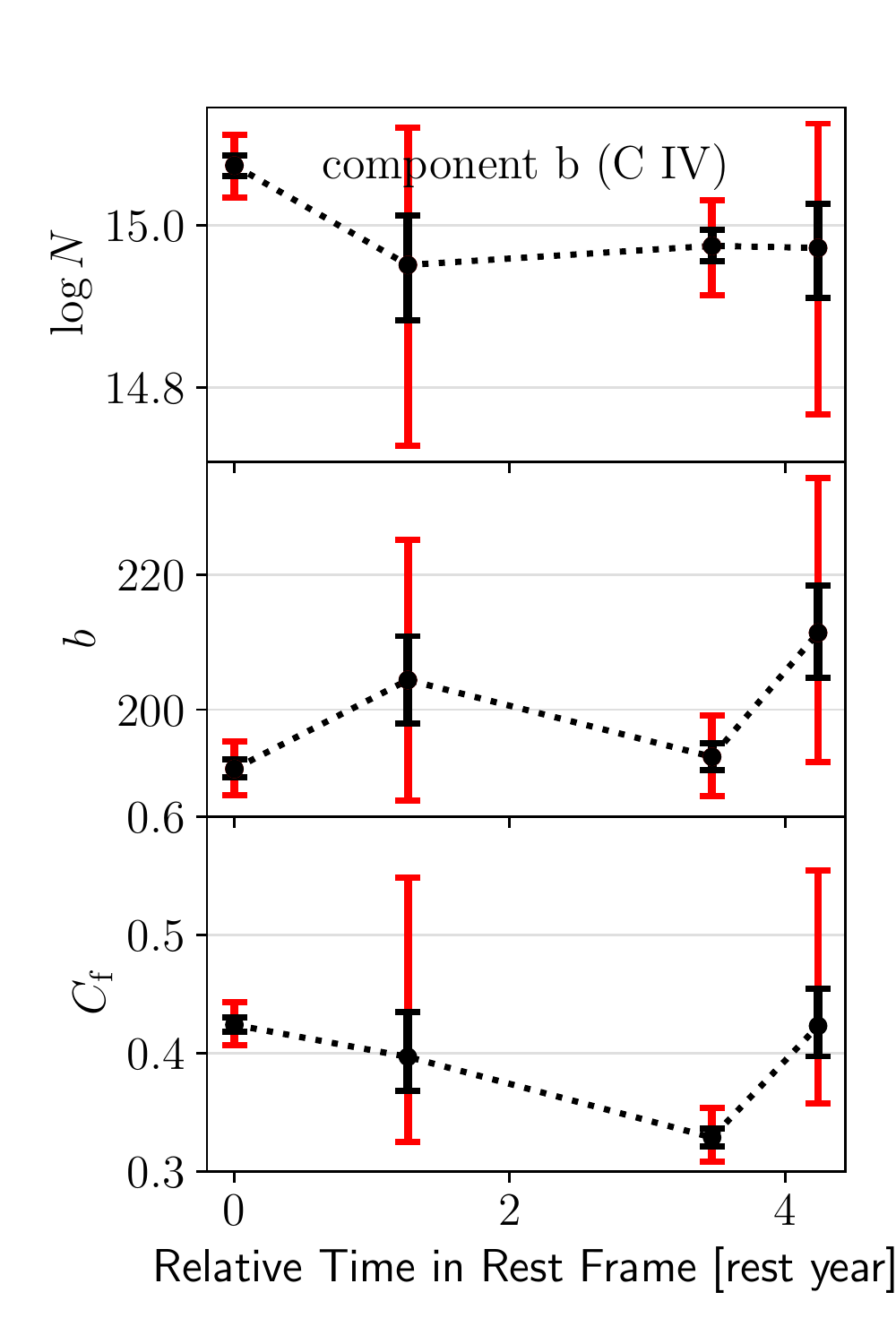}
    \includegraphics[width=8cm,angle=0]{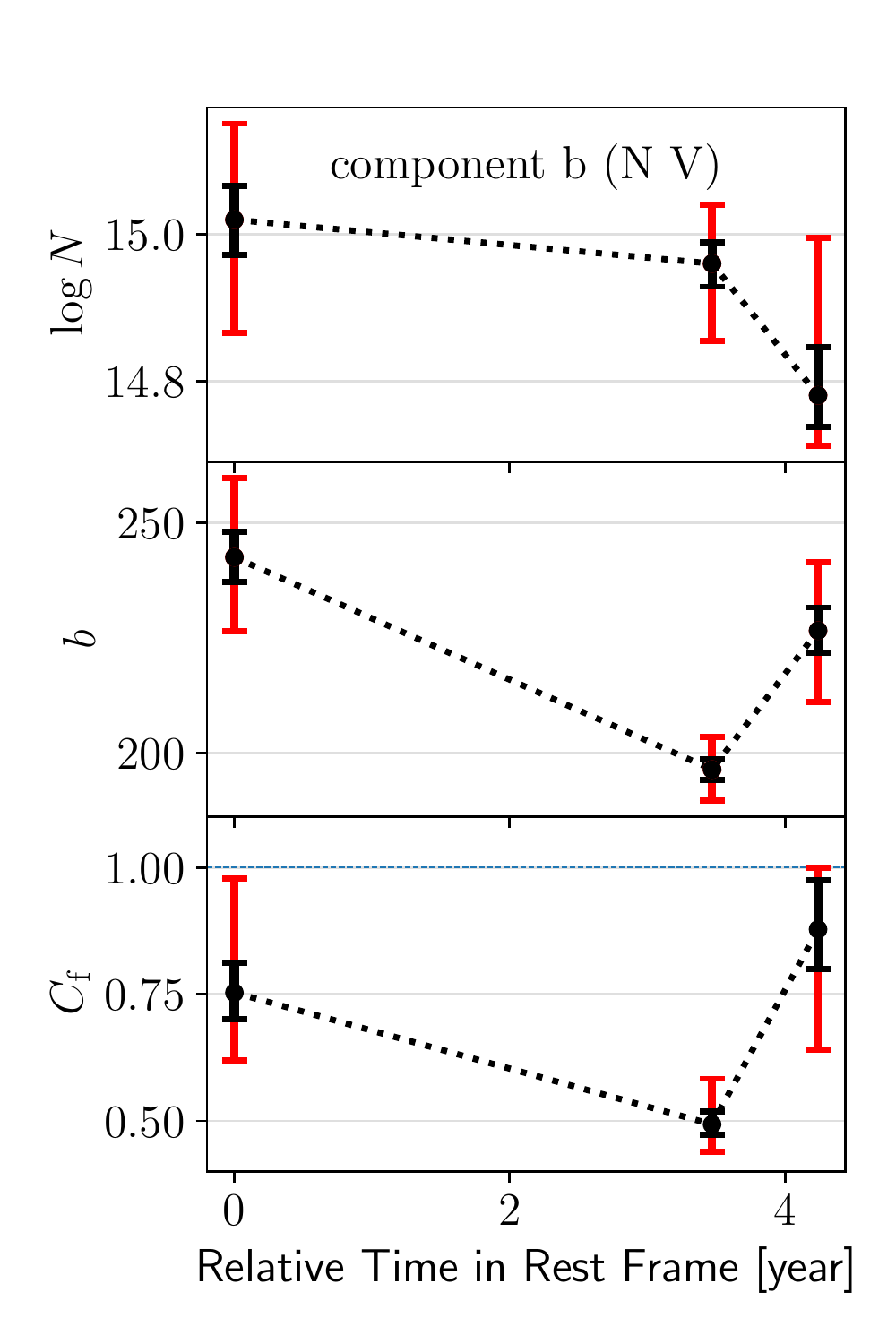}
 \end{center}
 \caption{Variation of fit parameters with time (\logN, $b$, and
   \cf\ from the top to the bottom) of component~b in \ion{C}{4}
   (left) and \ion{N}{5} (right) mini-BALs. Horizontal axis gives
   relative time from the first observing epoch (i.e., epoch~E1) in
   the quasar rest-frame.  Black dots show the mode of fit parameter
   posterior distributions, while black and red error bars denote
   1$\sigma$ and 3$\sigma$ uncertainties. Only \cf\ shows $>3\sigma$
   variability in \ion{C}{4} mini-BAL, while \cf\ and $b$ vary in the
   \ion{N}{5} mini-BAL.\label{fig:timevar}}
\end{figure*}

We find significant variability at $>$3$\sigma$ confidence level in
\cf\ of component~b both in the \ion{C}{4} and the \ion{N}{5}
mini-BALs as well as in $b$ of component~b only in the \ion{N}{5}
mini-BAL (see Table~\ref{tab:fit} and Figure~\ref{fig:timevar}) In
contrast, all parameters of components~n1 and n2 are stable over all
epochs at the $3\sigma$ level.

\begin{figure}
  \begin{center}
    \includegraphics[width=9cm,angle=0]{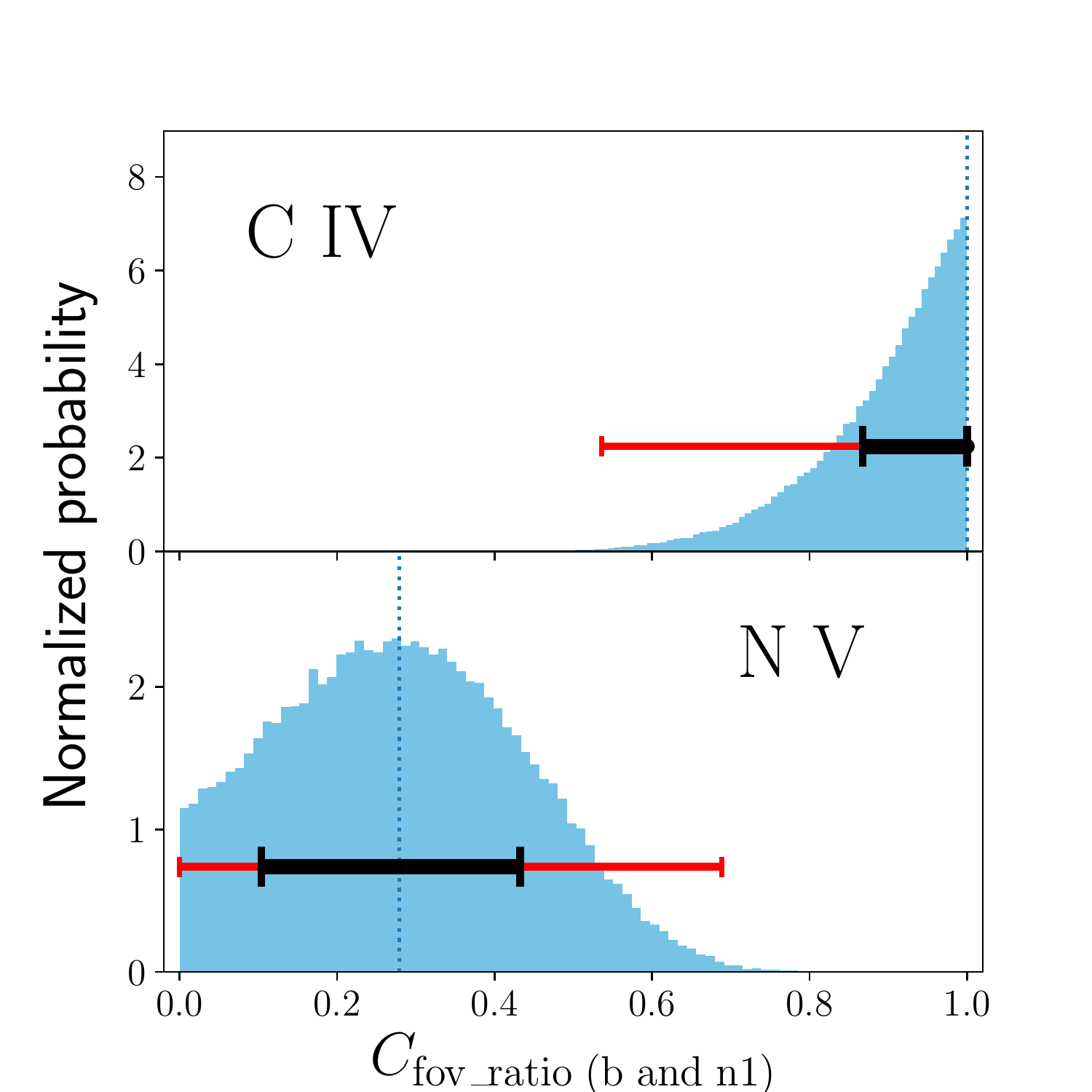}
 \end{center}
  \caption{Probability distribution histogram of \cfratio\ between
    components~b and n1 in epoch~E3 for \ion{C}{4} (top) and
    \ion{N}{5} (bottom). Vertical dotted lines are modes of the
    $C_{\rm fov\_ratio}$. Horizontal black and red error bars denote
    1$\sigma$ and 3$\sigma$ uncertainties.\label{fig:cfov}}
\end{figure}

We also study the overlap covering factor ratio (\cfratio) between
components~b and n1/n2.  We placed the strongest constraints on
\cfratio\ in the E3 spectrum.  As shown in Figure~\ref{fig:cfov},
components~b (\cf\ = 0.33$\pm$0.01) and n1 (\cf\ =
0.52$^{+0.02}_{-0.04}$) in the \ion{C}{4} mini-BAL are found to
overlap by at least 50\%\ of the projected size of the smaller
component (i.e., component~b) since the $3\sigma$ {\it lower} limit on
\cfratio\ is $\sim$0.54.  Thus, for the first time we find with high
confidence that multiple absorbing clouds along our line of sight
overlap with each other.  However, there need not be overlap between b
and n1 for \ion{N}{5}, since we found a $3\sigma$ {\it upper} limit on
\cfratio\ of $\sim$0.69 for the \ion{N}{5} mini-BAL (see
Figure~\ref{fig:cfov}). That is, in the case of
Figure~\ref{fig:overlap} the two \ion{C}{4} absorbers should overlap
along our line of sight by at least 50\%\ of their size, while the two
\ion{N}{5} absorbers do not necessarily overlap.  Here, we should
emphasize that the \ion{C}{4} and \ion{N}{5} absorbers do not
necessarily have an identical overlap covering factor ratio because
they may represent different layers in the absorber (i.e., they have
different covering factors).

\subsection{System~B} \label{subsec:sysB}
We detect only a weak \ion{C}{4} NAL with $EW_{\rm rest} \sim
0.03$~\AA\ in the system. The covering factor is \cf\ = 1, which
suggests an intervening absorber. A weak absorption profile on the
left side of the blue component is not part of
\ion{C}{4}~$\lambda$1548 since we do not find any absorption features
in the \ion{C}{4}~$\lambda$1551 window.  Therefore, we ignore this
region when carrying out the fit. The best fit model is shown in
Figure~\ref{fig:bestfit}.

\subsection{System~C} \label{subsec:sysC}
There are five components in \ion{C}{4} and two in \ion{Si}{4} in this
system, of which some are completely separated from each other by
unabsorbed spectral regions. However, we regard these components as a
single NAL system since their overall velocity separation is
$<$~200~\kms\ from each other, following \citet{mis07a}. The best fit
model is presented in Figure~\ref{fig:bestfit}.  The fit to
component~1 of the \ion{Si}{4} NAL implies partial coverage
(\cf\ $\sim$ 0.69) with a 3$\sigma$ confidence level. However, this
result should be regarded with caution because (i) there exist narrow
spikes in both the blue and red members of the \ion{Si}{4} doublet
that could be due to data defects and (ii) the best-fit model appears
to overestimate the depth of red component which would underestimate
\cf.  The \cfratio\ between components~4 and 5 in the \ion{C}{4} NAL
is consistent with unity, which is reasonable since both components have
full coverage.

\subsection{System~D} \label{subsec:sysD}
We detect \ion{C}{4} and \ion{Si}{4} NALs in this system.  We fit
models with two components to the \ion{C}{4} NAL because of its
asymmetric profile, while a single component model is acceptable for
the \ion{Si}{4} NAL (see Figure~\ref{fig:bestfit}). All components
are consistent with full coverage at the 1$\sigma$ confidence level.
It is reasonable that \cfratio\ $\sim$ 1 between the two components of
the \ion{C}{4} NAL since both have full coverage.

\subsection{System~E} \label{subsec:sysE}
We detect only a \ion{C}{4} NAL that consists of two
components. Both components are consistent with full coverage at the
3$\sigma$ level.  There could exist a third very weak component at
$\Delta~v$ $\sim$~60~\kms\ in Figure~\ref{fig:bestfit}, but we
ignore it since it is not detected with $>$~5$\sigma$ confidence
(i.e., does not satisfy our detection criterion).

\subsection{System~F} \label{subsec:sysF}
A shallow and broad \ion{C}{4} NAL with a Doppler parameter of $b$
$\sim$ 23~\kms\ is detected in the system as shown in
Figure~\ref{fig:bestfit}.  The system is probably an intervening
absorber, since \cf\ $\sim$ 1 at the 3$\sigma$ confidence level.

\subsection{System~G} \label{subsec:sysG}
Only a \ion{C}{4} NAL is detected in this system. Although we fit a
model with a single component to this NAL, there could be two
components since the \ion{C}{4}~$\lambda$1548 profile is slightly
asymmetric. For multi-component fitting, we would need a higher S/N
spectrum.  The covering factor is consistent with full coverage.  The
best-fit models are presented in Figure~\ref{fig:bestfit}.

\section{Discussion} \label{sec:resdis}

\subsection{Possible Geometry}\label{sec:IPC}
Using the MCMC-based method for fitting the mini-BALs in the spectrum
of UM675, we find for the first time that multiple absorbing clouds
(i.e., components~b and n1) overlap along our line of sight to the
background flux source, which could not have been inferred with the
$\chi^2$-based method. Thus, the projected distribution of integrated
total column density from multiple absorbing clouds along the line of
sight is much more complex than previously thought.  By synthesizing
artificial spectra, \citet{sab05} studied how the column density that
we derive from the observed spectrum depends on inhomogeneous partial
coverage. They found that homogeneous and inhomogeneous models have
little difference (apparent optical depths within 50\%) as long as the
column density distribution does not contain spatially narrow peaks
(i.e., large enhancements in the optical depth over small coverage
areas).

The fits to the \ion{C}{4} and the \ion{N}{5} mini-BALs in epoch~E3
provide the strongest constraints on partial coverage among the four
epochs because of the high S/N in that spectrum; \cf~$_{\rm (b)}$ =
0.33$^{+0.01}_{-0.01}$, \cf~$_{\rm (n1)}$ = 0.52$^{+0.02}_{-0.04}$,
and \cf~$_{\rm (n2)}$ = 0.38$^{+0.04}_{-0.11}$ for components b, n1,
and n2 of the \ion{C}{4} mini-BAL, and \cf~$_{\rm (b)}$ =
0.49$^{+0.02}_{-0.02}$, \cf~$_{\rm (n1)}$ = 0.16$^{+0.02}_{-0.02}$,
and \cf~$_{\rm (n2)}$ = 0.83$^{+0.17}_{-0.60}$ for those of the
\ion{N}{5} mini-BAL, respectively.  All components (except for
component~n2 of the \ion{N}{5} mini-BAL) show partial coverage with a
confidence $> 3\sigma$. The width of component~b (FWHM $\sim$200~\kms)
is much larger than those of narrow components (FWHM
$\sim$10--30~\kms), which suggests that only component~b has a large
velocity dispersion possibly due to mechanisms such as internal
turbulence or continuous acceleration, but that its transverse scale
is comparable to those of the narrow components.

\begin{figure}
  \begin{center}
    \includegraphics[width=8.5cm,angle=0]{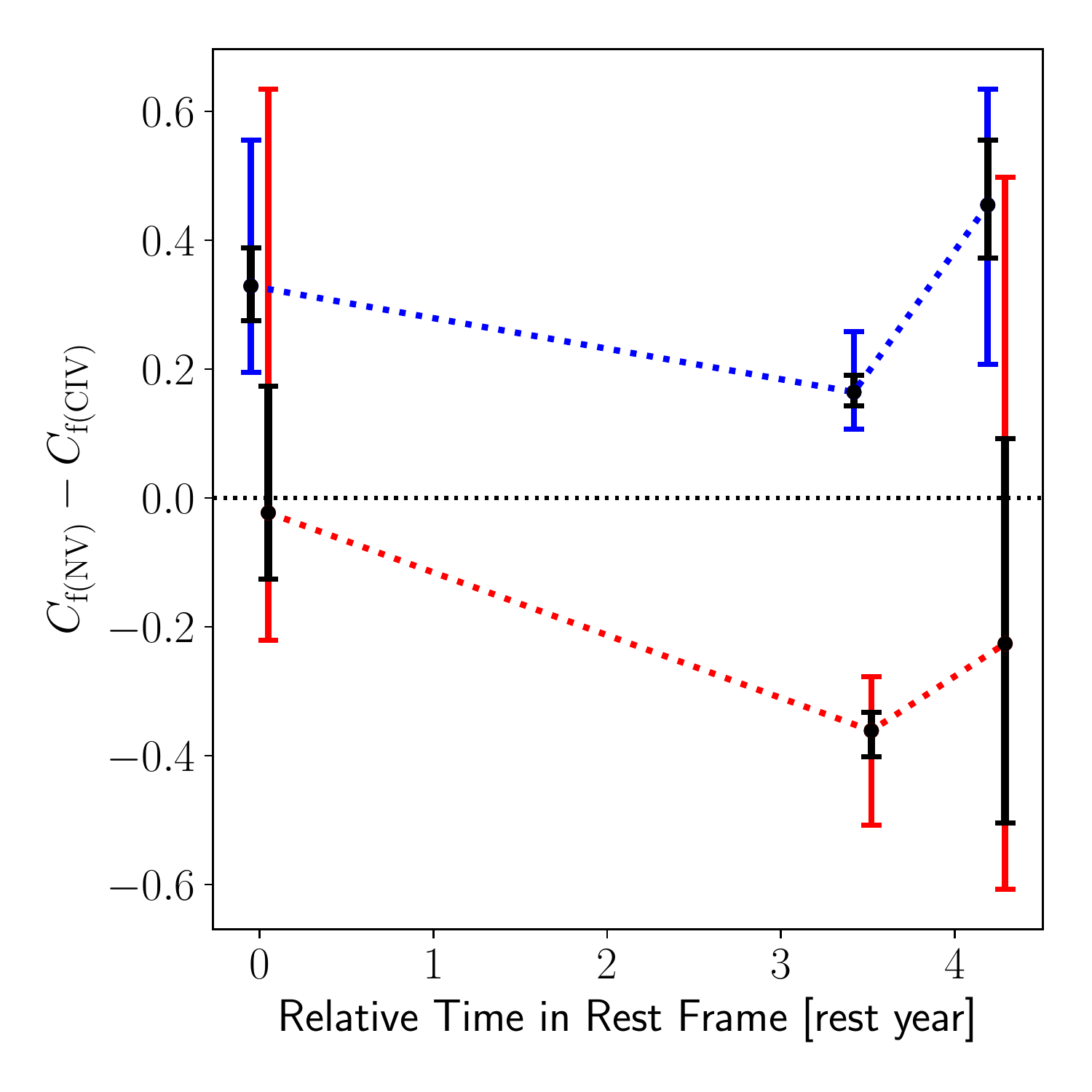}
 \end{center}
  \caption{Difference between \cf~(\ion{N}{5}) and \cf~(\ion{C}{4}) of
    components~b (blue) and n1 (red) as a function of relative time
    from the first epoch in the quasar rest frame.  Black dots are
    modes of the difference.  Black and blue/red error bars denote
    1$\sigma$ and 3$\sigma$ uncertainties of the difference. The dots
    with error bars are intentionally shifted by $\pm$0.05 years from
    the observed epoch in the horizontal direction for display
    purpose.\label{fig:cfcomp}}
\end{figure}

We also noticed that \cf~(\ion{N}{5}) is always larger than
\cf~(\ion{C}{4}) in all epochs for component~b. This trend (i.e.,
higher ions tend to have larger covering factors) is consistent with
other reports in the literature (e.g.,
\citealt{pet99,sri00,mis07a,muz16}), suggesting that the size of the
\ion{N}{5} absorbers is larger than that of the \ion{C}{4} absorbers
and/or the effective size of flux source behind the \ion{N}{5}
absorber is smaller than that behind the \ion{C}{4} absorber
(cf. Figure~\ref{fig:overlap}).  Interestingly, the opposite is true
  for components~n1 and n2; i.e., \cf~(\ion{N}{5}) tends to be {\it
    smaller} than \cf~(\ion{C}{4}) as shown in
  Figure~\ref{fig:cfcomp}. This discrepancy between the broad and the
  narrow components suggests that the size (or flux amplitude) of the
  background flux source is different between \ion{N}{5} and
  \ion{C}{4} absorbers.  This could happen for the narrow components
  if the \ion{N}{5} absorption line is more diluted than the
  \ion{C}{4} absorption line by the flux from the BELR whose scale
  ($\sim$0.1~pc; \citealt{ham97b} and references therein) is $\sim$2
  orders of magnitude larger than the size of the continuum source as
  we will discuss later.  Because both the \ion{C}{4} and \ion{N}{5}
  mini-BALs are located on the blue wing of the corresponding broad
  emission lines (see Figure~1 of \citealt{ham95}), their depths can
  be diluted by the broad emission line flux (e.g., \citealt{ara99}).

Considering contributions from both the continuum source and the BELR,
we can calculate the total covering factor \cf\ by
\begin{equation}
C_{\rm f} = \frac{C_{\rm f (cont)} + W \times C_{\rm f (BELR)}}{1 + W},
\end{equation}
where $C_{\rm f(cont)}$ and $C_{\rm f(BELR)}$ are covering factors of
the continuum source and BELR and $W$ (= $F_{\rm BELR} / F_{\rm c}$)
is defined as the ratio of the fluxes from the BELR ($F_{\rm BELR}$)
and the continuum source ($F_{\rm c}$) \citep{gan99}.  Indeed, the
flux ratio ($W$) around the \ion{N}{5} mini-BAL due to the
contribution from both \lya\ and \ion{N}{5} broad emission lines is
obviously larger than that around the \ion{C}{4} mini-BAL (see
Figure~1 of \citealt{ham95}).  Thus, the opposite trend of the ratio
of \cf~(\ion{C}{4}) and \cf~(\ion{N}{5}) in components~b and n1/n2 can
be due to a difference in the contribution from the background flux
source, as discussed in \citet{wu10}.

We speculate that component~b does not absorb light from the BELR (it
may be at smaller radial distance than the BELR, $r$ $\leq$ 0.1~pc or
embedded within it), while components~n1 and n2 also absorb light from
the BELR (possibly because they are located at a larger radial
distance); we refer to this scenario as model~A, hereafter.  The very
high ionization state of the gas in component~b, showing \ion{Ne}{8}
and \ion{O}{6}, also supports its small radial distance from the
continuum source.  In other words, components~b and n1/n2 are not
co-spatial but happen to be located along the same cylinder of sight
with a similar offset velocity perhaps due to
line-locking.\footnote{The absorption line is aligned with other lines
  in a spectrum because they share radiation pressure at the same
  wavelength. This is one of the clues that the QALs are radiatively
  accelerated in the vicinity of the quasars (e.g., \citealt{bow14}).}
Just as in the UM675 spectrum, narrow kinematic components are
sometimes detected near the centers of mini-BALs in other quasars
(e.g., HE1341$-$1020, Q1157$+$014, and Q2343$+$125) and they usually
show no variability while broader components vary
\citep{ham97a,mis14}.  Possible origins of these narrow components
include (i) dense clumpy clouds with large volume density (i.e., less
affected by fluctuation of incident flux) embedded in the mini-BAL
flow \citep{mis14} and (ii) gas clouds in the quasar host galaxy or in
foreground galaxies that are physically unrelated to the quasar
\citep{ham97a}.  In addition to these, we propose a third possible
origin: (iii) dense clumpy clouds with large volume density whose
radial distance is a few orders of magnitude larger than the mini-BAL
flow (e.g., \citealt{ito20}) but with their radial velocity
coinciding. The variety of absorption lines with a wide range of
ionization potential \citep{bea91,ham95} also supports this idea.

Although the above scenario is possible, it is unlikely that the broad
and narrow absorbers at very different distances have similar
velocities relative to the quasar by chance. In that spirit, we
present an alternative model, model~B, hereafter, in which the broad
and the narrow absorbers coexist at the same radial distance. In this
model, most of the absorber has relatively a low density and high
ionization state and large velocity gradient (corresponding to
component~b). Within the large gas parcel there are dense clumps with
low ionization state and small velocity gradient (corresponding to
components~n1 and n2). The entire absorber is moving both in the
transverse and radial directions as it orbits around the flux
sources. The size of component~b is comparable to the size of the
continuum source so that it can partially cover both the continuum
source and the BELR. With this geometry, components n1 and n2 can also
produce partial coverage since they are smaller than component~b.

In either model, the geometry of the mini-BAL system in UM675 requires
two distinct flux sources (i.e., the continuum source and BELR), while
{\tt mc2fit} (as well as other $\chi^2$-based codes) assumes a single
background flux source. To verify this geometry, we need to introduce
covering factors (\cf) and their overlap covering factor ratios
(\cfratio) as free parameters for both background flux sources
individually, or introduce a model for the brightness distribution of
the background flux source(s) and then adopt specific geometrical
arrangements for the absorbers, or both. However, the current version
of {\tt mc2fit} does not have these features. These are improvements
that we will introduce in a future version of the code.

One possible observational test to distinguish between models~A and B
is through monitoring of the outflow velocity of the narrow
components. In model~B, components~n1 and n2 should show a velocity
shift on a time scale similar to the dynamical (i.e., crossing) time
of the BELR, $t_{\rm dyn}$ since they are located at a radial distance
comparable to the size of the BELR. Assuming a Keplerian motion, we
derive $t_{\rm dyn}$ $\sim$23~years in the quasar's rest frame which
is much longer than our current monitoring period of $\sim$4.24~years.

\subsection{Origin of Time Variability}\label{sec:timevar}
In addition to the variability in absorption strength
\citep{ham95,mis14}, we also find variability in the line parameters
of the mini-BAL system: component~b shows $>3\sigma$ variability in
\cf\ of the \ion{C}{4} mini-BAL and in both \cf\ and $b$ of the
\ion{N}{5} mini-BAL in $\sim$4.24~years in the quasar's rest frame.
Since the time separation between epochs~E3 and E4 is the smallest
($\Delta t_{\rm rest}$ $\sim$0.77~year) among the four observing
epochs, we concentrate on the variability pattern on that short time
interval since it can place the most stringent constraints on the
physical conditions of the absorbers.

There are two major scenarios for the origin of time variability that
have often been discussed in the literature: (1) a change of the
ionization state of the gas clouds (e.g.,
\citealt{mis07b,ham11,fil13,hor16}) and (2) motion of the absorbing
clouds across our line of sight to the background flux source (e.g.,
\citealt{gib08,ham08,viv16,kro17}).  Neither situation is applicable
to intervening absorbers, unless they have very large gas density
and/or sharp edges \citep{nar04}.

The column densities of component~b in the \ion{C}{4} and \ion{N}{5}
mini-BALs are almost constant while their \cf\ values vary (see
Figure~\ref{fig:timevar}), which supports the gas motion scenario.  To
examine this scenario, we first need to estimate the sizes of the
background flux sources (i.e., the continuum source and
BELR). Following \citet{mis05}, we take five times the gravitational
radius as the continuum source size using a black hole mass of
$\log(M_{\rm BH}/M_\odot)$ = 9.52 \citep{hor16}. For the size of the
portion of the BELR that emits the \ion{C}{4} line, we use
equation~(1) of \citet{lir18} with Galactic extinction of $A_V$ = 0.44
and assume a power-law index of $\alpha$ = 0.61 \citep{lus15}.  We
obtain $R_{\rm cont}$ $\sim$ $2\times10^{-3}$~pc and $R_{\rm BELR}$
$\sim$ $0.2$~pc as the sizes of the continuum source and the BELR,
respectively.

Since component~b of the \ion{C}{4} mini-BAL shows variability in
covering factor from \cf\ = 0.33$^{+0.01}_{-0.01}$ to
0.42$^{+0.03}_{-0.02}$ in $\Delta t_{\rm rest}$ = 0.77~years, we can
place a lower limit on the crossing velocity\footnote{Here, we assume
  that a homogeneous absorber moves with a constant crossing
  velocity.}  as $v_{\rm cross} \geq d_{\rm s} \times \Delta C_{\rm f}
/\Delta t_{\rm rest}$, where $d_{\rm s}$ is the size of the continuum
source $d_{\rm cont}$ (or the size of the absorber $d_{\rm abs}$) if
$d_{\rm abs}$ is larger (or smaller) than $d_{\rm cont}$ respectively.
We adopt $\Delta C_{\rm f} = 0.09$ as the variability amplitude of
\cf, and take $\Delta t_{\rm rest}$ to be the time interval between
observing epochs in the quasar's rest frame.  Since we want to place a
lower limit on $v_{\rm cross}$, we should use a smaller size between
$d_{\rm abs}$ and $d_{\rm cont}$ (i.e., $d_{\rm s}$ = min\{$d_{\rm
  abs}, d_{\rm cont}$\}).  However, these sizes should be almost the
same since the absorber shows partial coverage, which leads us to use
$d_{\rm s}$ $\sim$ $d_{\rm cont}$ whose value we have been able to
estimate.  Substituting the values above into $d_{\rm cont} \times
\Delta C_{\rm f} /\Delta t_{\rm rest}$, we obtain $v_{\rm cross}$
$\geq$~366~\kms.  If the crossing velocity is equal to the Keplerian
orbital velocity around the central black hole\footnote{The enclosed
  mass is much larger than the black hole mass if the absorber's
  radial distance is as large as the size of the host galaxy.}, we can
also place a weak constraint on the absorber's radial distance from
the flux source as $\leq$~106~pc for component~b.  We also note that
in the context of model~A this radial distance is probably smaller
than the size of BELR ($<$ $R_{\rm BELR}$ $\sim$ 0.2~pc) as discussed
above.

In comparison, components~n1 and n2 do not show significant (i.e.,
$>3\sigma$) variability during our monitoring campaign in $\Delta
t_{\rm rest}$ = 4.24~years.  Therefore, we cannot place any
constraints on either crossing velocity or radial distance.

\section{Summary} \label{sec:sum}
In this study, we introduce a Bayesian approach combined with an MCMC
method for fitting a mini-BAL and six NAL systems in the spectrum of a
radio-loud quasar UM675 taken with Keck/HIRES and Subaru/HDS in four
epochs. Our methodology is implemented in a new code, {\tt mc2fit},
which we test here using synthetic and real spectra. Our main results
are as follows:

\begin{itemize}
\item{Using {\tt mc2fit}, we restrict the range of covering factor
  (\cf) from 0 to 1 as is physically possible and determine the fit
  parameters more accurately than the traditional $\chi^2$-based
  methods.  Our fitting results for synthetic spectra are fairly
  consistent with those obtained by the $\chi^2$-based method
  described in \citet{mis07a}, but the latter method slightly
  underestimate \logN\ and $b$ compared to the correct values.}
\item{In addition to the original fit parameters (i.e., \zabs, \logN,
  $b$, \cf), we introduce the overlap covering factor ratio by
  multiple absorbers along our line of sight (\cfratio). Because of
  the large number of fit parameters ($\left[\sum_{i=2}^{n} {n \choose
      i}\right] + 4n$ in total for $n$ components), any observed
  spectrum to which we apply {\tt mc2fit} must have a very high S/N.
  Otherwise, the code tends to overestimate \cfratio\ especially when
  the correct value is small.}
\item{Among seven absorption systems (Systems~A -- G) in the quasar
  UM675, only the \ion{C}{4} and \ion{N}{5} mini-BALs at \zabs\ $\sim$
  2.1341 (System~A) show clear simultaneous variability in their
  \cf\ and $b$ at $>3\sigma$ confidence level.  We also determine for
  the first time that multiple absorbing clouds (i.e., the broad and
  the narrow components) overlap each other along our line of sight.}
\item{For component~b of the mini-BAL system (System~A),
  \cf~(\ion{N}{5}) is always larger than \cf~(\ion{C}{4}) at all
  epochs, while the opposite trend holds for components~n1 and n2.
  These trends suggest that the broad component does not absorb light
  from the BELR or that the narrow components do not absorb light from
  the continuum source, depending on the possible models we have
  considered.}
\item{The column densities of component~b in the \ion{C}{4} and
  \ion{N}{5} mini-BALs are almost constant while their \cf\ values
  vary, which supports the gas motion scenario for their variability.
  If this is the case, the broad component must be at $r<106$~pc
  (and could be closer than the size of $R_{\rm BELR}$, $\sim$
  0.2~pc) with a rotational velocity of $v_{\rm rot} >
  366$~\kms\ assuming Keplerian motion (we cannot place any
  meaningful constraints for the narrow components).}
\end{itemize}  

The MCMC-based approach improves fitting results significantly
compared to the traditional $\chi^2$-based methods especially for
high-quality spectra with S/N $\geq$ 30~pixel$^{-1}$. It also provides
the overlap covering factor ratio (\cfratio) if multiple components
overlap in the spectrum, which conveys important information on the
absorber's geometry. More detailed information can be obtained if we
introduce fit parameters for the continuum source and BELR,
separately.  To capitalize on the MCMC-based technique, we plan to
monitor several mini-BAL systems whose absorption profiles are easy to
de-blend using high quality spectra like the one in UM675.

\acknowledgments
We are honored and grateful for the opportunity of observing the
Universe from Maunakea, which has the cultural, historical and natural
significance in Hawaii.  We would like to thank the anonymous referee
for very useful comments and suggestions.  We also would like to thank
Fred Hamann for providing us with his Keck/HIRES data, Christopher
Churchill for providing us with the {\tt MINFIT} software package, and
Takashi Horiuchi for valuable comments. This work was supported by
JSPS KAKENHI Grant Number 21H01126.

\section*{Appendix}
Best-fit models for absorption lines in Systems~B -- G using the MCMC
method. The velocity on the horizontal axis is defined relative to
the flux-weighted center of the absorption profiles.  The top and
middle panels show the profiles of the blue and red members of a
doublet on a common velocity scale, with the model profile produced by
{\tt mc2fit} superposed: dashed lines for each component and a solid
line for the contribution of all components.  The bottom panel shows
the two profiles together, along with the resulting covering factors
(open circles) with their 1$\sigma$ error bars. Profiles in the shaded
areas are ignored for fitting.

\addtocounter{figure}{-6}
\begin{figure}[h]
  \begin{center}   
    \includegraphics[width=9cm,angle=0]{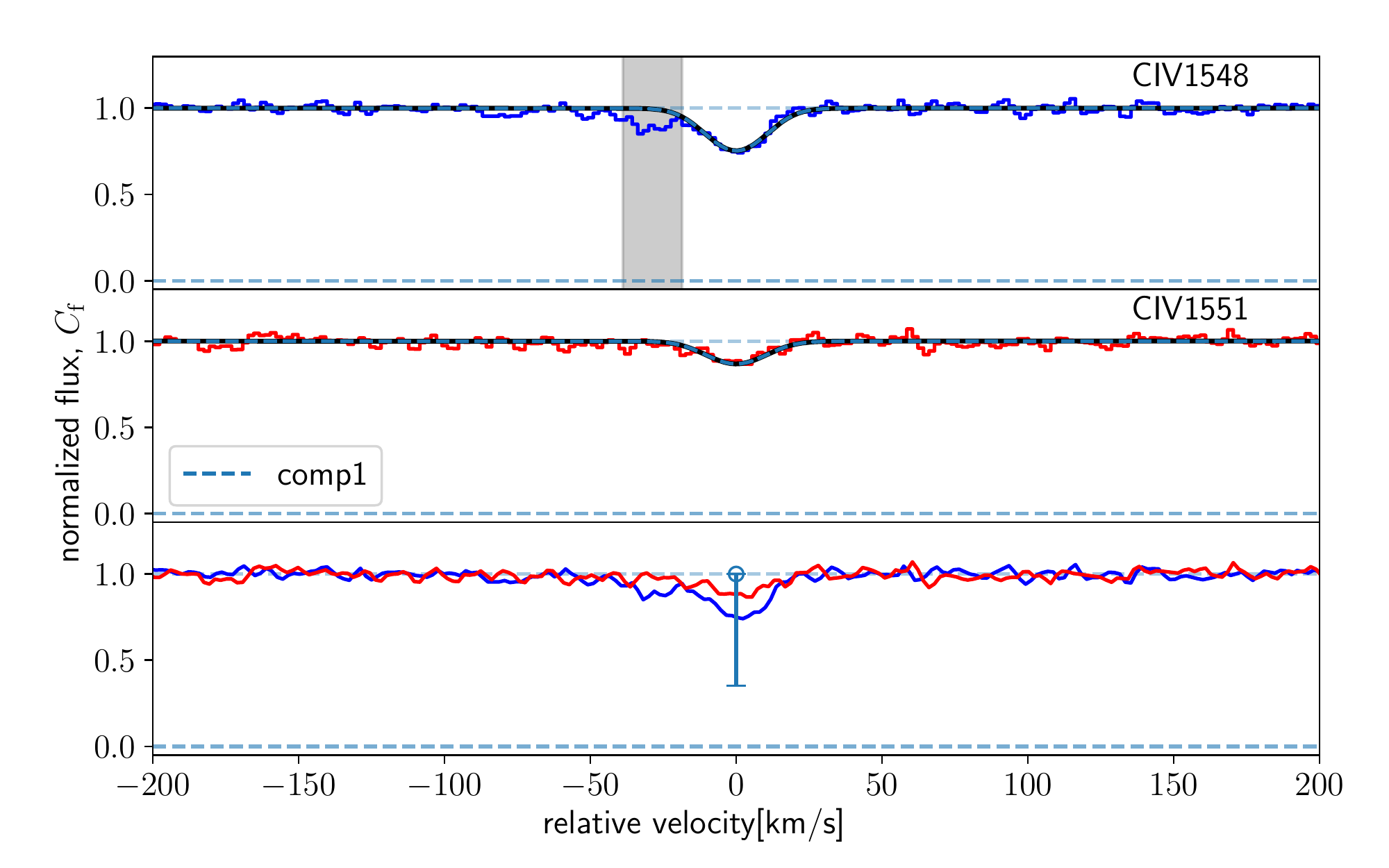}
  \end{center}
  \caption{Best-fit model for the \ion{C}{4} NAL in System~B using
    the MCMC method.}
\end{figure}

\addtocounter{figure}{-1}
\begin{figure}[h]
  \begin{center}   
    \includegraphics[width=9cm,angle=0]{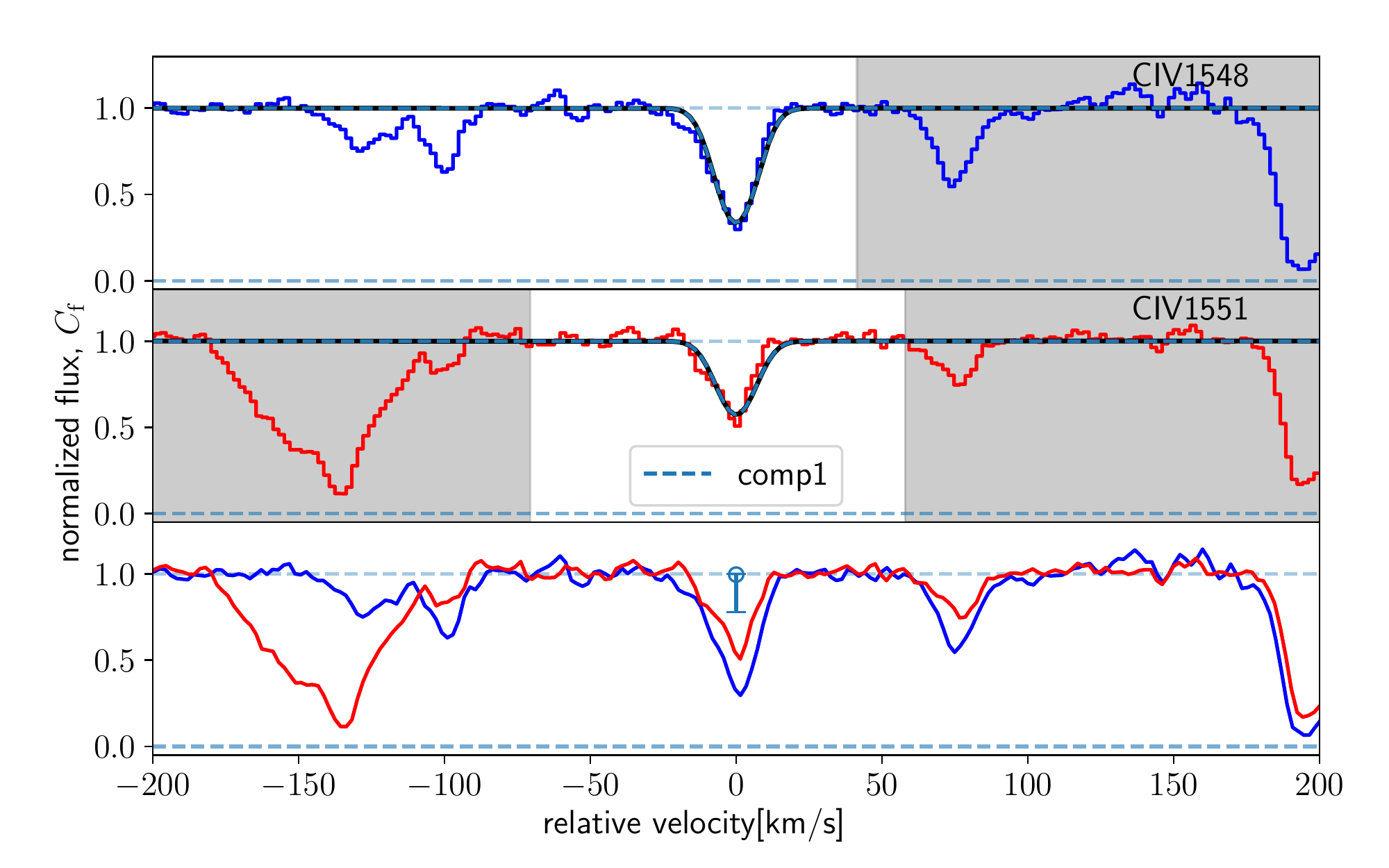}
    \includegraphics[width=9cm,angle=0]{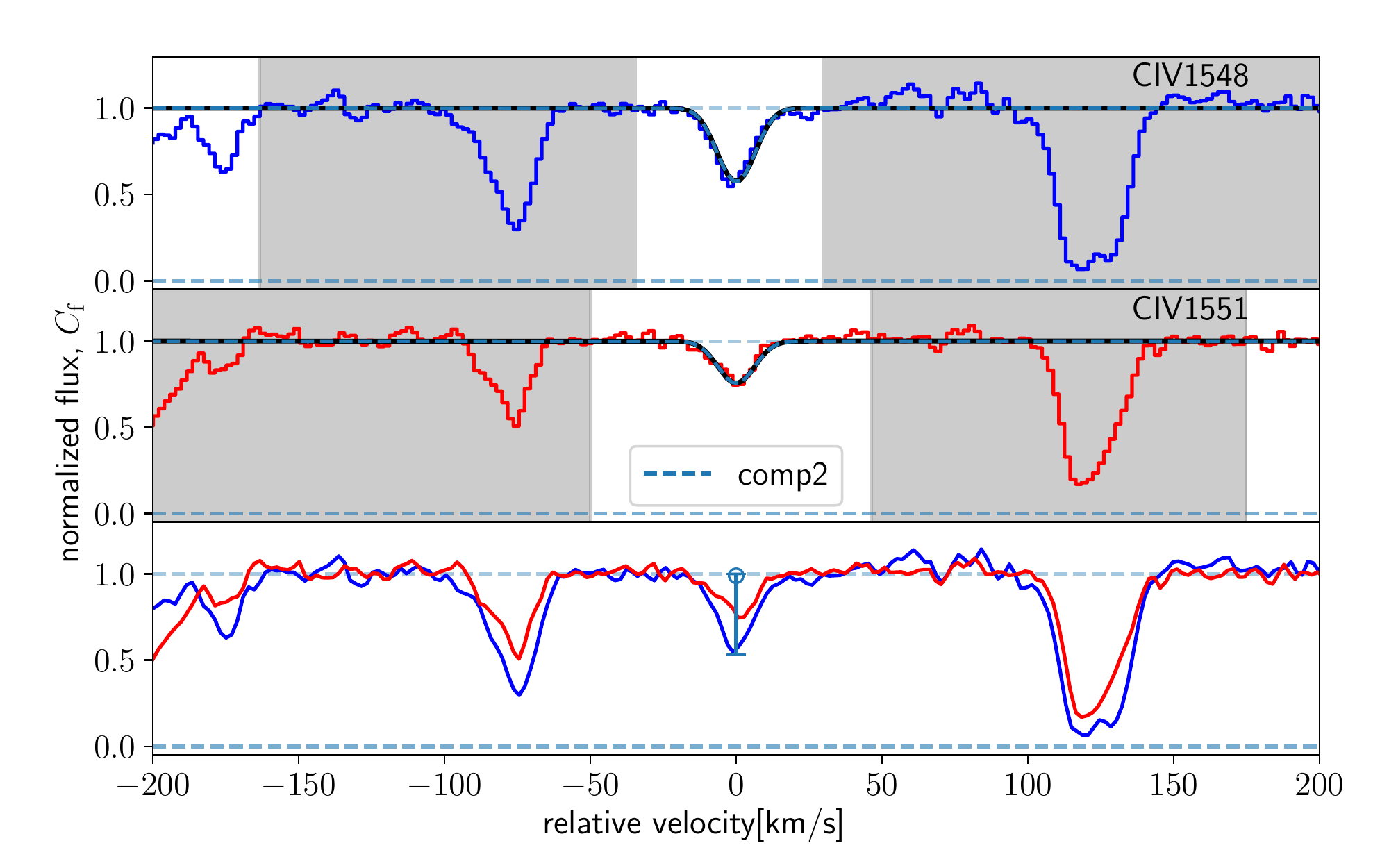}
    \includegraphics[width=9cm,angle=0]{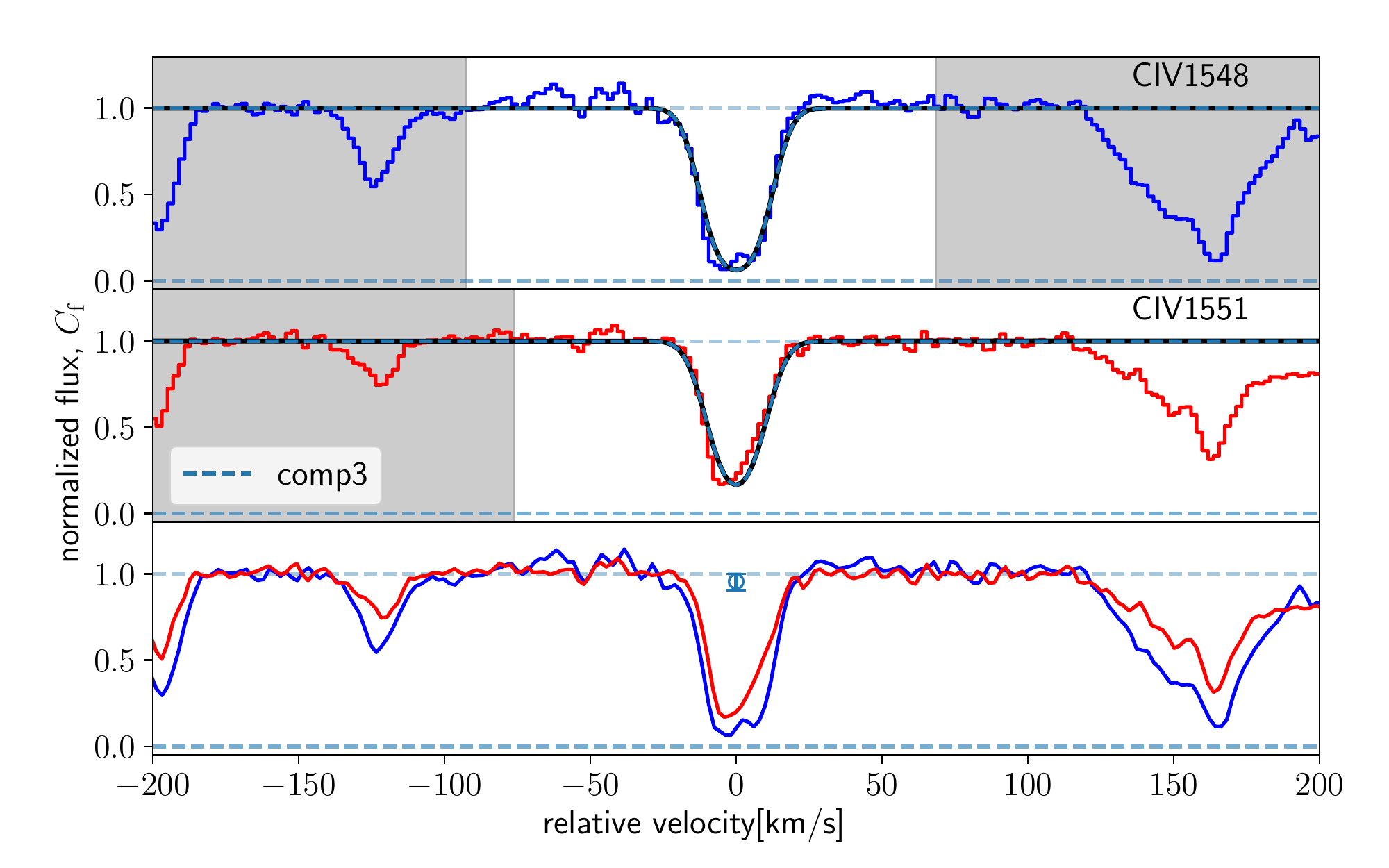}
  \end{center}
  \caption{Best-fit models for the \ion{C}{4} and \ion{Si}{4} NALs in
    System~C using the MCMC method.}
\end{figure}

\addtocounter{figure}{-1}
\begin{figure}[h]
  \begin{center}   
    \includegraphics[width=9cm,angle=0]{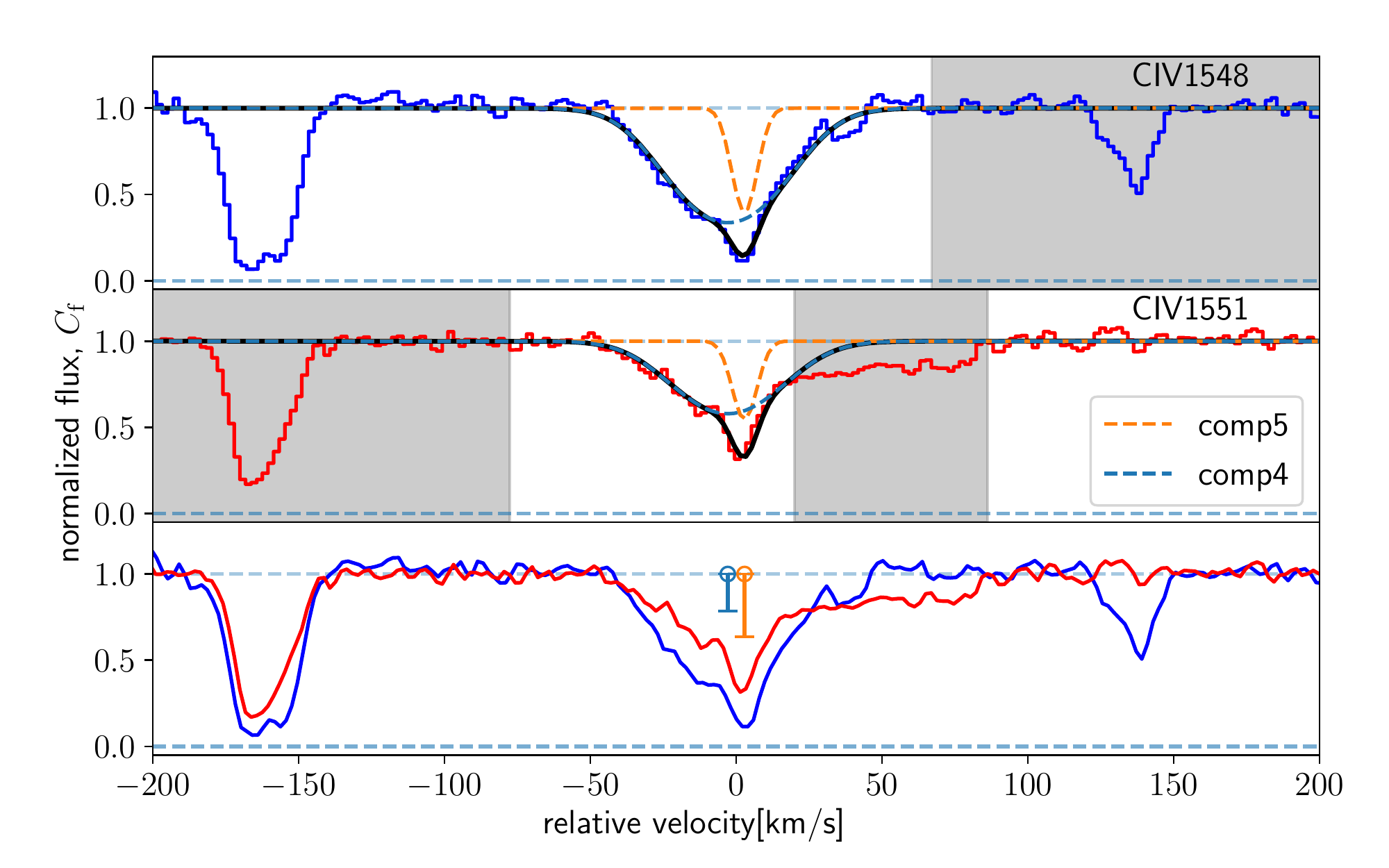}
    \includegraphics[width=9cm,angle=0]{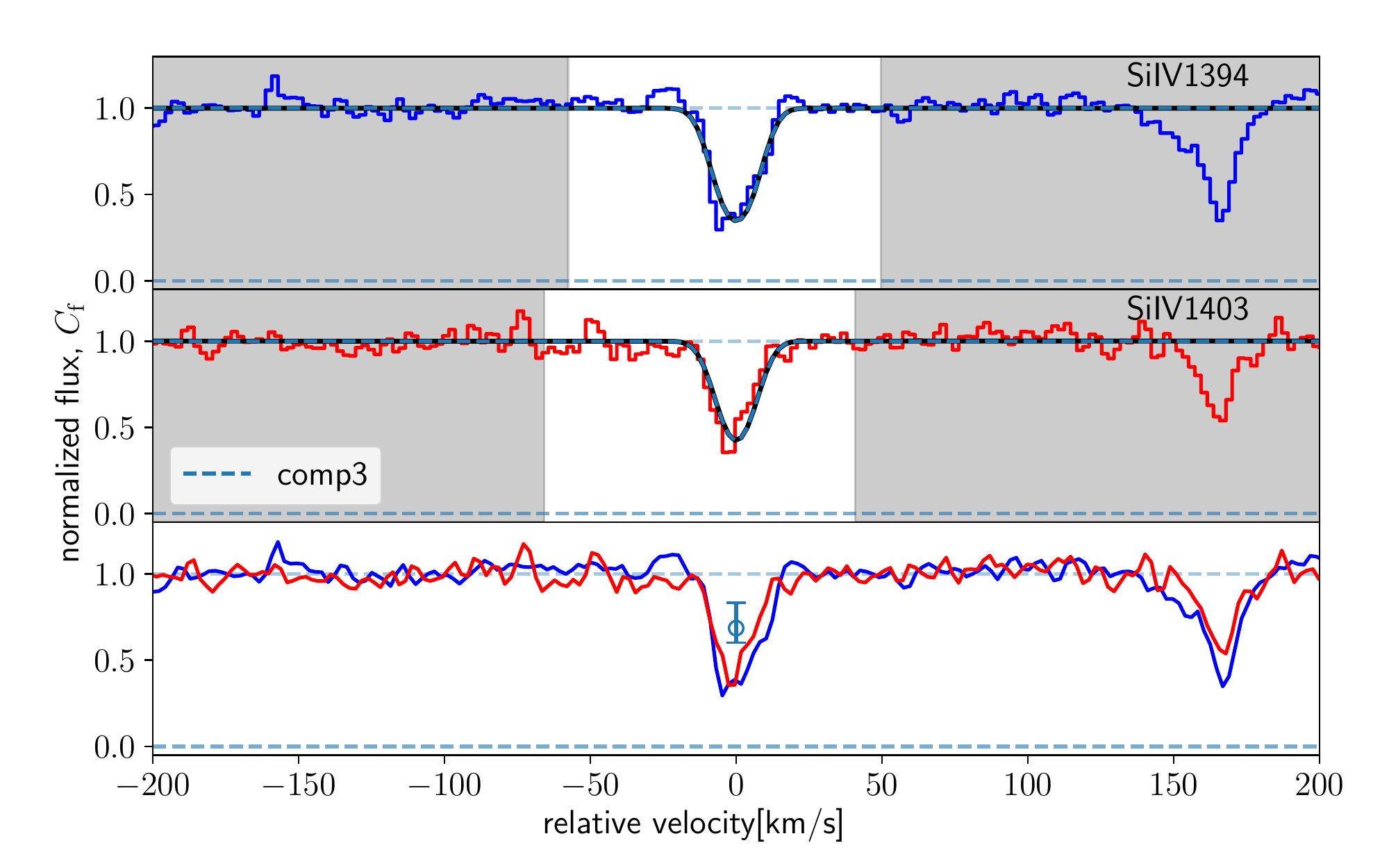}
    \includegraphics[width=9cm,angle=0]{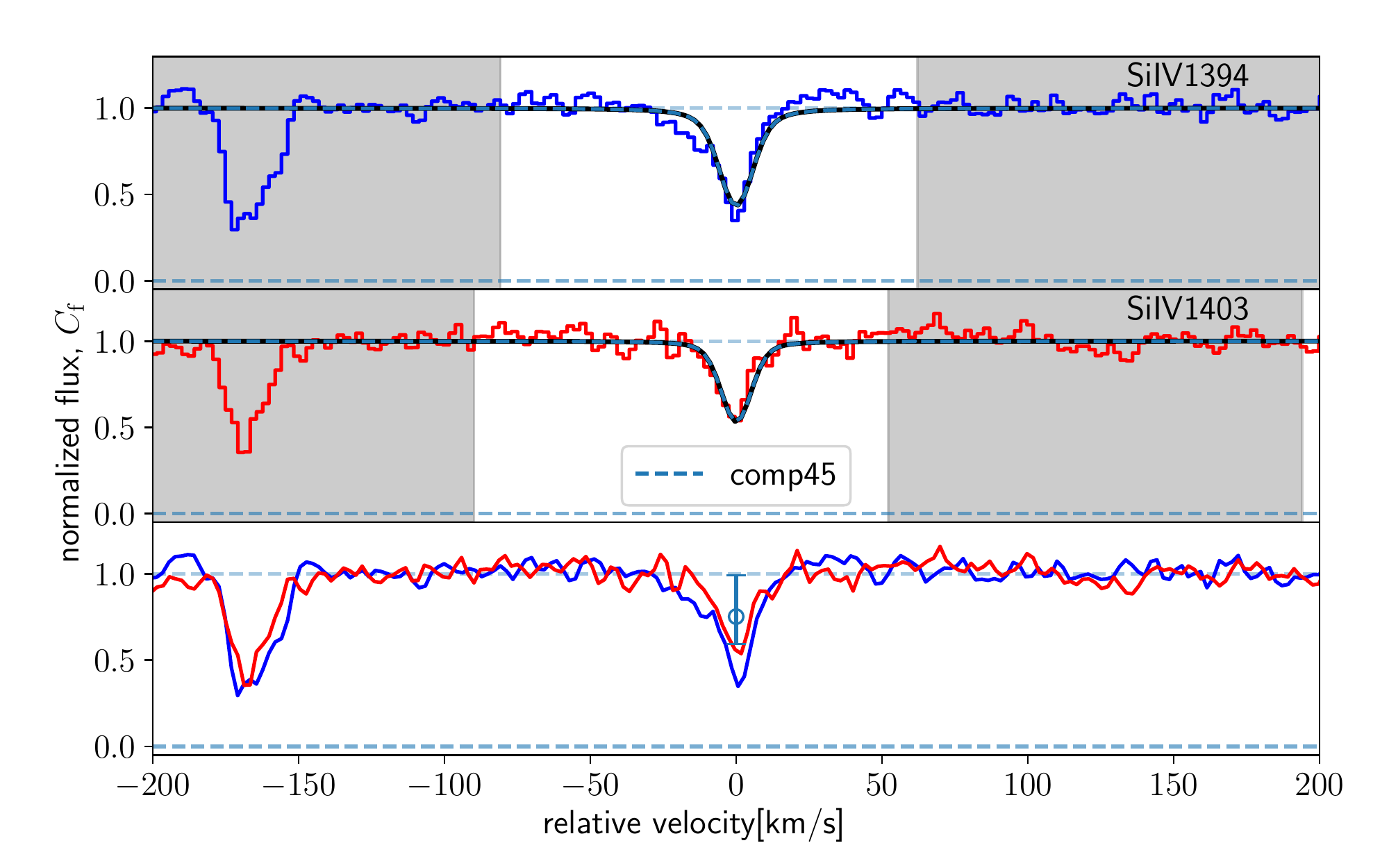}
  \end{center}
  \caption{Best-fit models for the \ion{C}{4} and \ion{Si}{4} NALs in
    System~C using the MCMC method.}
\end{figure}

\addtocounter{figure}{-1}
\begin{figure}[h]
  \begin{center}
    \includegraphics[width=9cm,angle=0]{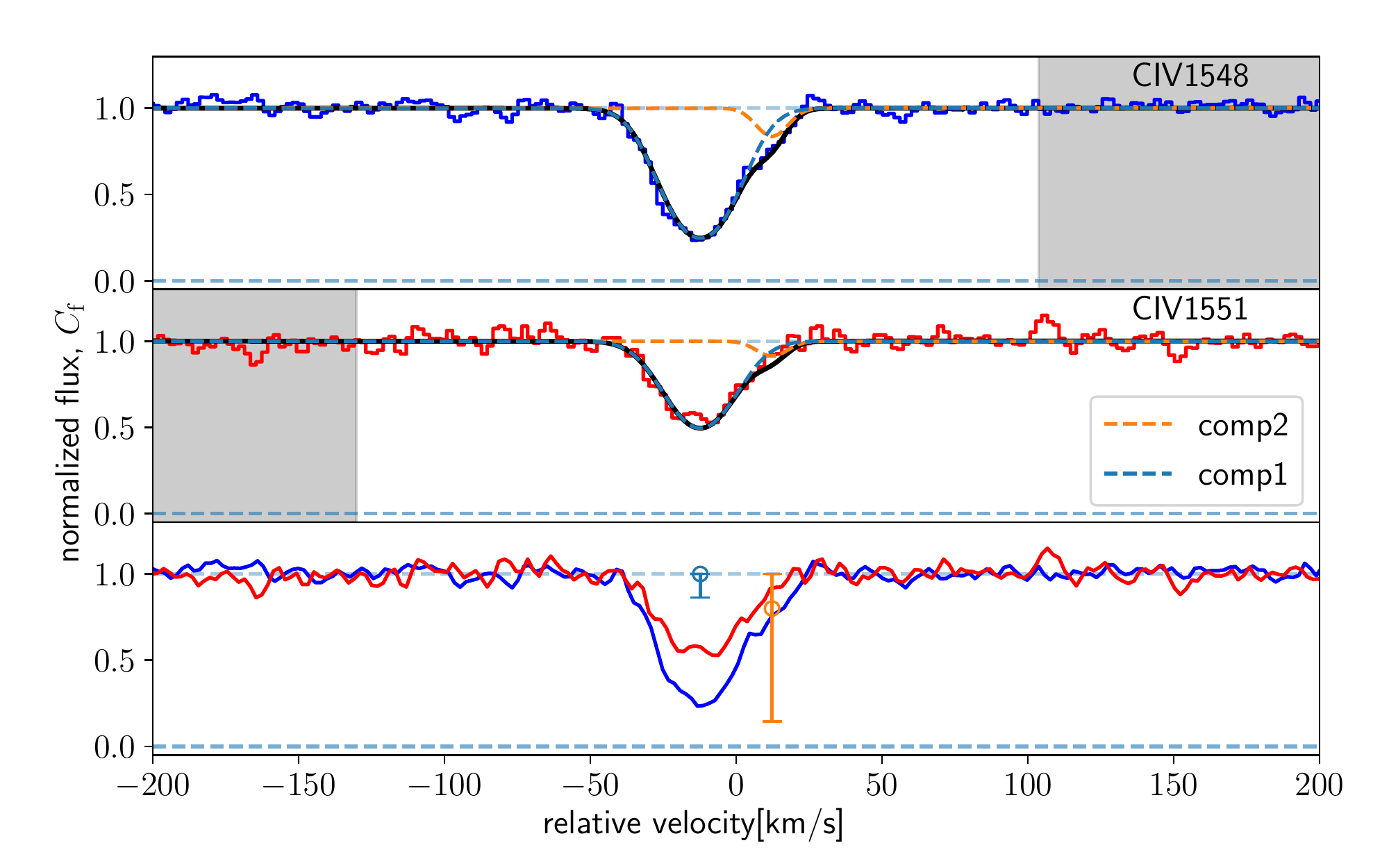}
    \includegraphics[width=9cm,angle=0]{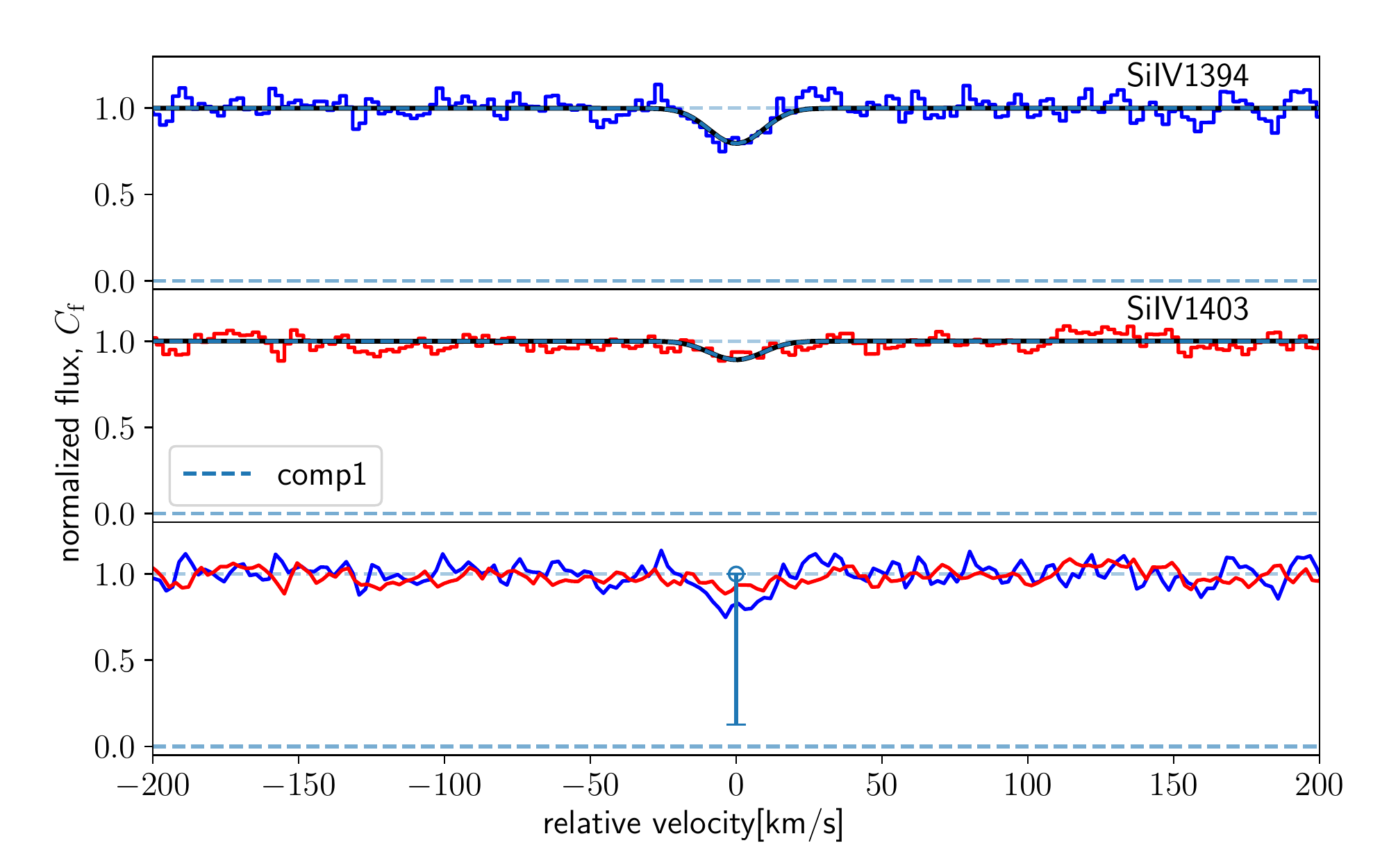}
  \end{center}
  \caption{Best-fit models for the \ion{C}{4} and \ion{Si}{4} NALs in
    System~D using the MCMC method.}
\end{figure}

\addtocounter{figure}{-1}
\begin{figure}[h]
  \begin{center}
    \includegraphics[width=9cm,angle=0]{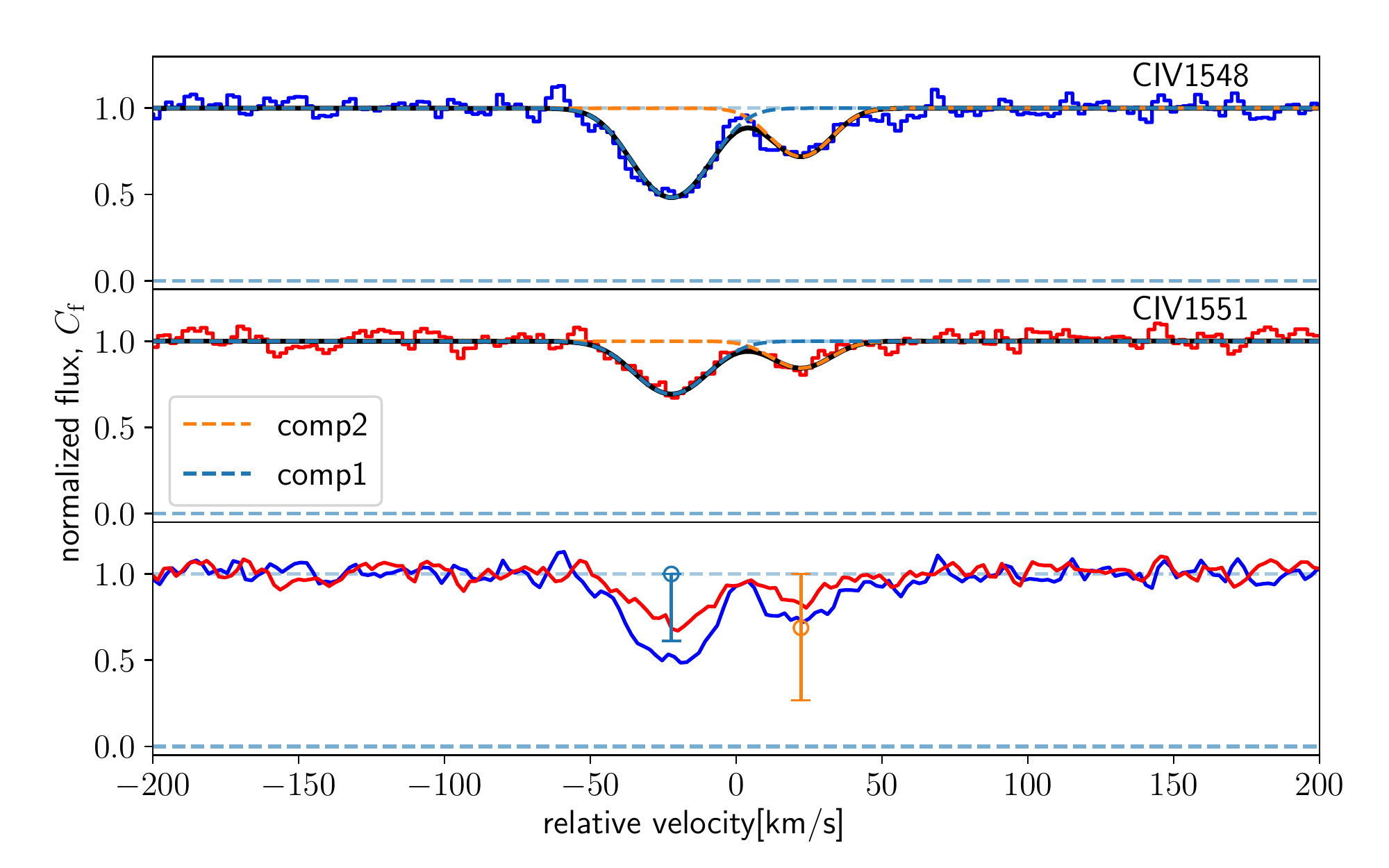}
  \end{center}
  \caption{Best-fit model for the \ion{C}{4} NAL in System~E using the
    MCMC method.}
\end{figure}

\addtocounter{figure}{-1}
\begin{figure}[h]
  \begin{center}
    \includegraphics[width=9cm,angle=0]{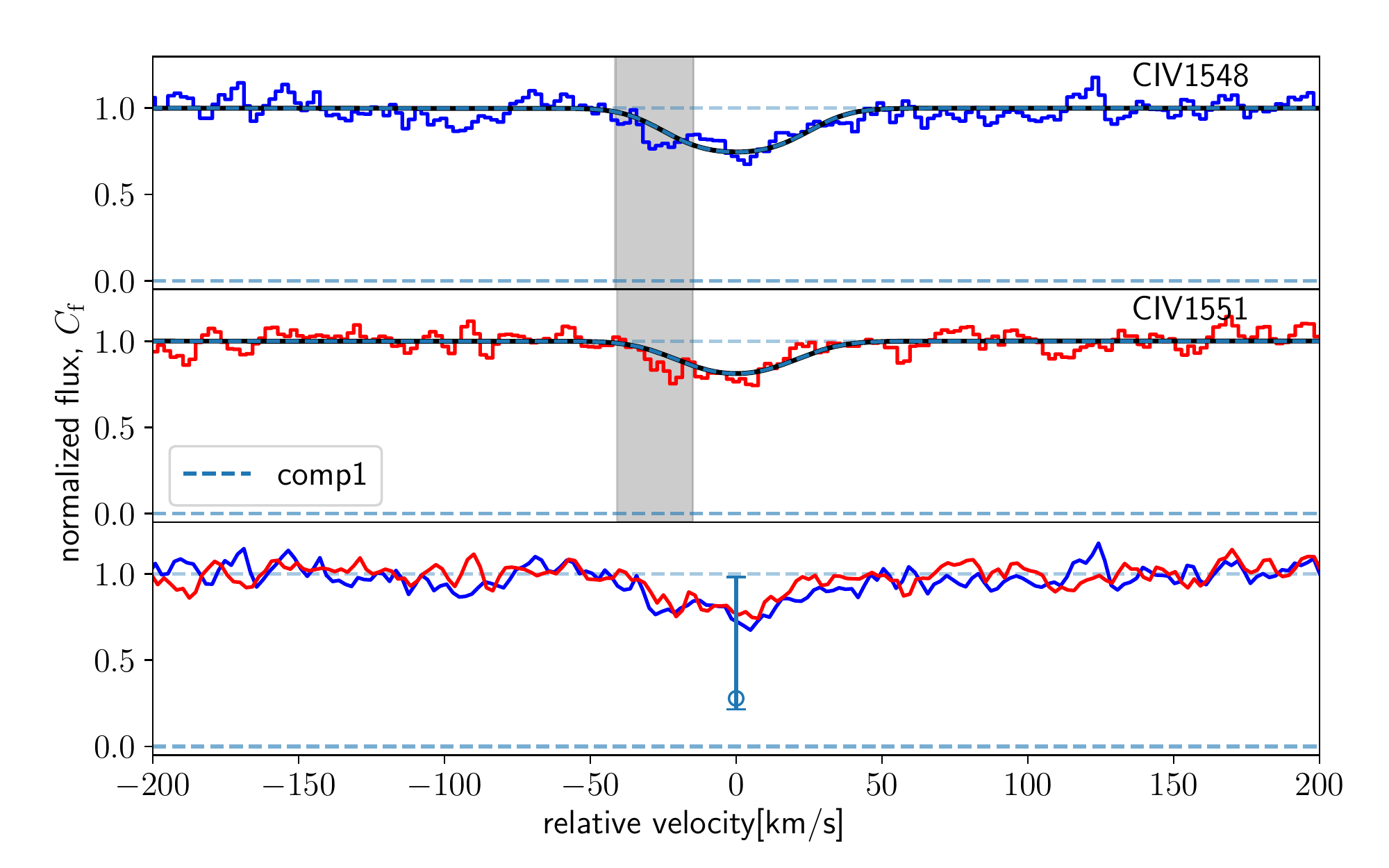}
  \end{center}
  \caption{Best-fit model for the \ion{C}{4} NAL in System~F using the
    MCMC method.}
\end{figure}

\addtocounter{figure}{-1}
\begin{figure}[h]
  \begin{center}
    \includegraphics[width=9cm,angle=0]{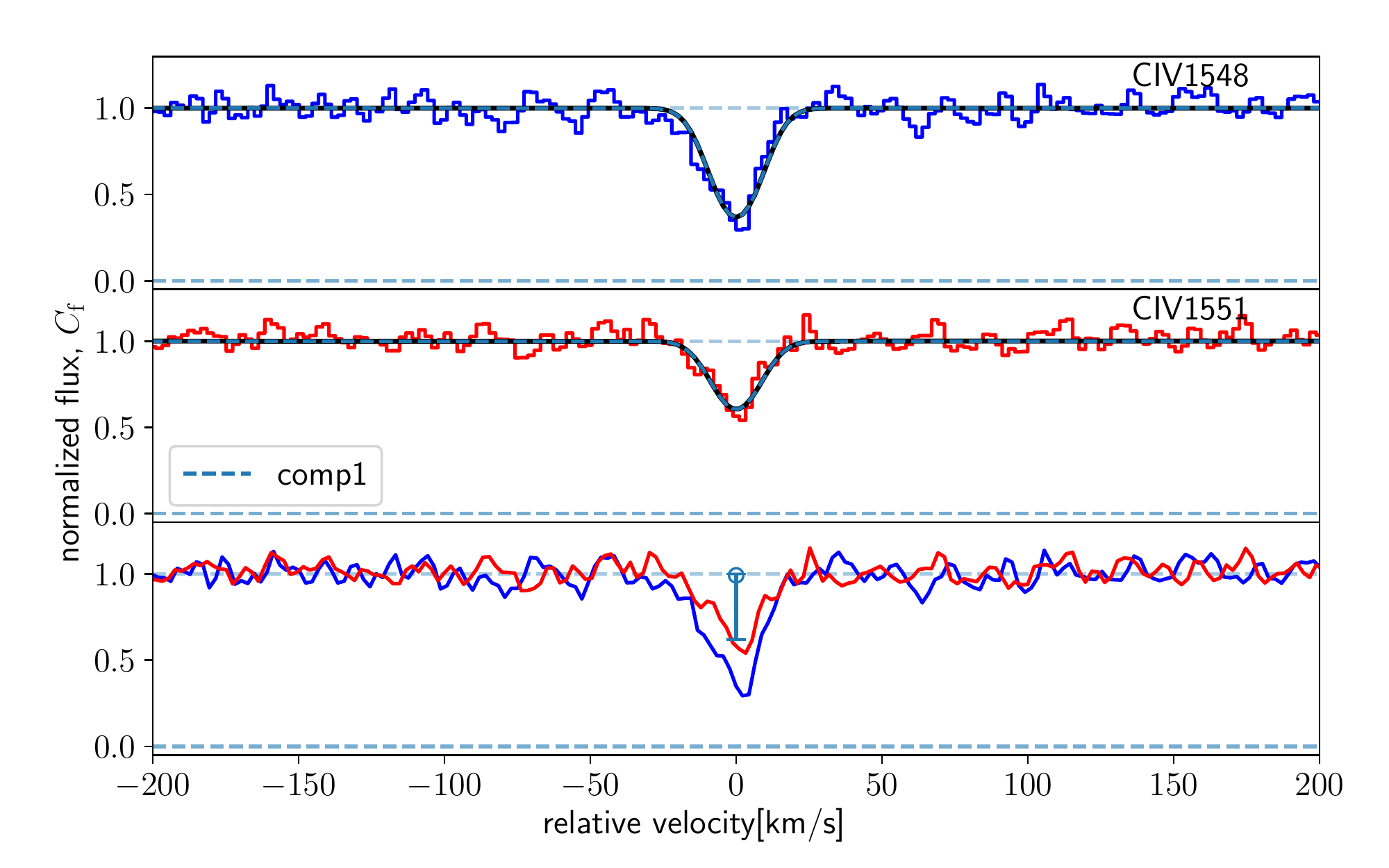}
  \end{center}
  \caption{Best-fit model for the \ion{C}{4} NAL in System~G using the
    MCMC method.}
\end{figure}

\clearpage
\LongTables
\begin{deluxetable*}{ccccccccc}
\tablecaption{Fitting Parameters of mini-BAL System in All Epochs\label{tab:fit}}
\tablewidth{0pt}
\tablehead{
\colhead{Ion}             &
\colhead{Epoch}           &
\colhead{comp.}           &
\colhead{\zabs}           &
\colhead{$\log{N}^a$}     &
\colhead{$b^a$}           &
\colhead{\cf$^a$}         &
\colhead{\cfratio$^{a,b}$} &
\colhead{EW$_{\rm rest}^c$}  \\
\colhead{}                &
\colhead{}                &
\colhead{}                &
\colhead{}                &
\colhead{(\cmm)}          &
\colhead{(\kms)}          &
\colhead{}                &
\colhead{}                &
\colhead{(\AA)}           \\
}
\startdata
\multicolumn{9}{c}{{\rm System~A}} \\
\hline
\ion{C}{4}  & 1 & b  & 2.1344 & $15.07_{-0.01(-0.04)}^{+0.01(+0.04)}$ & $191.3_{-1.4(-4.0)}^{+1.3(+4.0)}$  & $0.42_{-0.01(-0.02)}^{+0.01(+0.02)}$ &                               & $1.72 \pm 0.07$ \\
            &   & n1 & 2.1341 & $14.12_{-0.15(-0.33)}^{+0.09(+0.21)}$ &  $28.2_{-1.3(-3.7)}^{+1.4(+3.8)}$  & $0.31_{-0.08(-0.12)}^{+0.18(+0.43)}$ & $0.40_{-0.25(-0.39)}^{+0.24(+0.60)}$ &                 \\
            &   & n2 & 2.1334 & $13.62_{-0.11(-0.31)}^{+0.22(+0.69)}$ &  $12.7_{-1.2(-3.9)}^{+1.1(+3.2)}$  & $0.52_{-0.23(-0.40)}^{+0.05(+0.14)}$ & $0.65_{-0.11(-0.62)}^{+0.35(+0.35)}$ &                 \\
            & 2 & b  & 2.1347 & $14.95_{-0.07(-0.23)}^{+0.06(+0.18)}$ & $204.3_{-6.3(-18.0)}^{+6.7(+21.0)}$ & $0.40_{-0.03(-0.07)}^{+0.04(+0.16)}$ &                              & $1.47 \pm 0.17$ \\
            &   & n1 & 2.1341 & $13.96_{-0.26(-0.37)}^{+0.15(+0.47)}$ &  $30.9_{-2.5(-7.1)}^{+2.8(+8.6)}$  & $0.66_{-0.18(-0.39)}^{+0.16(+0.34)}$ & $1.00_{-0.36(-0.98)}^{+0.00(+0.00)}$ &                 \\
            & 3 & b  & 2.1343 & $14.98_{-0.02(-0.06)}^{+0.02(+0.06)}$ & $193.0_{-1.9(-5.8)}^{+2.0(+6.2)}$  & $0.33_{-0.01(-0.02)}^{+0.01(+0.03)}$ &                               & $1.22 \pm 0.06$ \\
            &   & n1 & 2.1341 & $14.13_{-0.03(-0.10)}^{+0.04(+0.13)}$ &  $25.6_{-0.6(-1.9)}^{+0.6(+1.8)}$  & $0.52_{-0.04(-0.14)}^{+0.02(+0.04)}$ & $1.00_{-0.13(-0.46)}^{+0.00(+0.00)}$ &                 \\
            &   & n2 & 2.1334 & $13.58_{-0.13(-0.40)}^{+0.15(+0.95)}$ &  $10.3_{-1.2(-4.7)}^{+1.3(+4.3)}$  & $0.38_{-0.11(-0.28)}^{+0.04(+0.11)}$ & $1.00_{-0.30(-0.91)}^{+0.00(+0.00)}$ &                 \\
            & 4 & b  & 2.1345 & $14.97_{-0.06(-0.20)}^{+0.06(+0.16)}$ & $211.6_{-6.8(-19.5)}^{+6.8(+22.5)}$ & $0.42_{-0.02(-0.06)}^{+0.03(+0.13)}$ &                               & $1.57 \pm 0.30$ \\
            &   & n1 & 2.1341 & $14.34_{-0.14(-0.34)}^{+0.31(+1.61)}$ &  $32.9_{-6.2(-16.4)}^{+4.2(+13.4)}$ & $0.51_{-0.27(-0.36)}^{+0.03(+0.11)}$ & $0.69_{-0.13(-0.67)}^{+0.31(+0.31)}$ &                 \\
\ion{N}{5}  & 1 & b  & 2.1343 & $15.02_{-0.05(-0.16)}^{+0.04(+0.13)}$ & $242.6_{-5.5(-16.0)}^{+5.4(+17.3)}$ & $0.75_{-0.05(-0.13)}^{+0.06(+0.23)}$ &                               & $1.96 \pm 0.33$ \\
            &   & n1 & 2.1341 & $14.81_{-0.40(-0.90)}^{+0.55(+1.19)}$ &  $17.3_{-4.2(-7.9)}^{+5.6(+17.4)}$  & $0.28_{-0.07(-0.17)}^{+0.07(+0.50)}$ & $1.00_{-0.21(-0.93)}^{+0.00(+0.00)}$ &                \\
            &   & n2 & 2.1334 & $13.53_{-0.18(-0.30)}^{+0.44(+2.45)}$ &   $9.9_{-2.9(-7.4)}^{+2.9(+9.4)}$  & $0.24_{-0.06(-0.10)}^{+0.67(+0.76)}$ & $1.00_{-0.39(-1.00)}^{+0.00(+0.00)}$ &                 \\
            & 3 & b  & 2.1343 & $14.96_{-0.03(-0.10)}^{+0.03(+0.08)}$ & $196.5_{-2.3(-6.7)}^{+2.4(+7.2)}$  & $0.49_{-0.02(-0.06)}^{+0.02(+0.09)}$ &                               & $1.05 \pm 0.11$ \\
            &   & n1 & 2.1341 & $15.48_{-0.25(-0.69)}^{+0.37(+0.52)}$ &  $15.1_{-0.8(-1.9)}^{+2.2(+6.8)}$  & $0.16_{-0.02(-0.03)}^{+0.02(+0.07)}$ & $0.28_{-0.18(-0.28)}^{+0.15(+0.41)}$ &                 \\
            &   & n2 & 2.1334 & $12.98_{-0.11(-0.23)}^{+0.32(+1.44)}$ &   $8.6_{-1.3(-4.2)}^{+1.4(+4.4)}$  & $0.83_{-0.60(-0.72)}^{+0.17(+0.17)}$ & $1.00_{-0.87(-1.00)}^{+0.00(+0.00)}$ &                 \\
            & 4 & b  & 2.1342 & $14.78_{-0.04(-0.07)}^{+0.07(+0.22)}$ & $226.8_{-5.0(-15.9)}^{+4.8(+14.5)}$ & $0.88_{-0.07(-0.24)}^{+0.10(+0.12)}$ &                               & $1.45 \pm 0.49$ \\
            &   & n1 & 2.1340 & $14.00_{-0.27(-0.48)}^{+0.36(+1.88)}$ &  $24.3_{-4.1(-14.2)}^{+3.9(+11.6)}$ & $0.29_{-0.09(-0.14)}^{+0.32(+0.71)}$ & $1.00_{-0.16(-1.00)}^{+0.00(+0.00)}$ &                 
\enddata
\tablenotetext{a}{1$\sigma$ upper/lower error bars in
  superscript/subscript. 3$\sigma$ uncertainties are in parenthesis.}
\tablenotetext{b}{Overlap covering factor ratio between
  components~b and n1/n2.}
\tablenotetext{c}{Total rest-frame equivalent width including all
  components. The 1$\sigma$ error is calculated as the combination of
  contributions from uncertainties in the intensities of individual
  pixels in the spectrum and from uncertainties in the placement of
  the continuum \citep{mis14}.}
\end{deluxetable*}

\clearpage
\LongTables
\begin{deluxetable}{cccccccc}
\tablecaption{Fitting Parameters of NAL Systems in epoch~E1\label{tab:fit2}}
\tablewidth{0pt}
\tablehead{
\colhead{Ion}             &
\colhead{comp.}           &
\colhead{\zabs}           &
\colhead{$\log{N}^a$}     &
\colhead{$b^a$}           &
\colhead{\cf$^a$}         &
\colhead{\cfratio$^{a,b}$} &
\colhead{EW$_{\rm b\_rest}^c$} \\
\colhead{}                &
\colhead{}                &
\colhead{}                &
\colhead{(\cmm)}          &
\colhead{(\kms)}          &
\colhead{}                &
\colhead{}                &
\colhead{(\AA)}           \\
}
\startdata
\multicolumn{8}{c}{{\rm System~B}} \\
\hline
\ion{C}{4}  & 1 & 2.0569 & $13.00_{-0.04(-0.09)}^{+0.15(+0.52)}$ & $13.2_{-0.9(-2.6)}^{+1.0(+3.3)}$ & $1.00_{-0.30(-0.65)}^{+0.00(+0.00)}$ &                               & $0.03 \pm 0.01$ \\
\hline
\multicolumn{8}{c}{{\rm System~C}} \\
\hline
\ion{C}{4}  & 1 & 2.0063 & $13.39_{-0.02(-0.05)}^{+0.05(+0.18)}$ &  $7.5_{-0.3(-0.9)}^{+0.3(+0.9)}$ & $1.00_{-0.08(-0.22)}^{+0.00(+0.00)}$ &                               & $0.06 \pm 0.00$ \\
            & 2 & 2.0071 & $13.08_{-0.03(-0.08)}^{+0.11(+0.38)}$ &  $6.8_{-0.5(-1.5)}^{+0.5(+1.6)}$ & $0.99_{-0.19(-0.46)}^{+0.01(+0.01)}$ &                               & $0.04 \pm 0.00$ \\
            & 3 & 2.0083 & $14.00_{-0.03(-0.08)}^{+0.03(+0.11)}$ &  $9.0_{-0.2(-0.6)}^{+0.2(+0.6)}$ & $0.96_{-0.02(-0.05)}^{+0.02(+0.04)}$ &                               & $0.13 \pm 0.01$ \\
            & 4 & 2.0099 & $13.80_{-0.02(-0.04)}^{+0.04(+0.14)}$ & $23.6_{-0.5(-1.6)}^{+0.5(+1.7)}$ & $1.00_{-0.07(-0.22)}^{+0.00(+0.00)}$ &                               & $0.17 \pm 0.01^d$ \\
            & 5 & 2.0100 & $13.32_{-0.08(-0.19)}^{+0.12(+2.27)}$ &  $2.7_{-0.6(-1.7)}^{+0.6(+1.6)}$ & $1.00_{-0.08(-0.36)}^{+0.00(+0.00)}$ & $1.00_{-0.02(-0.22)}^{+0.00(+0.00)}$ &                 \\
\ion{Si}{4} & 1$^e$ & 2.0083 & $13.49_{-0.13(-0.34)}^{+0.18(+1.33)}$ &  $5.9_{-0.6(-2.4)}^{+0.5(+1.4)}$ & $0.69_{-0.03(-0.08)}^{+0.04(+0.15)}$ &                               & $0.05 \pm 0.01$ \\
            & 2 & 2.0099 & $13.04_{-0.12(-0.22)}^{+0.11(+0.36)}$ &  $6.3_{-0.7(-1.9)}^{+0.7(+2.1)}$ & $0.74_{-0.08(-0.18)}^{+0.13(+0.26)}$ &                               & $0.04 \pm 0.01$ \\
\hline
\multicolumn{8}{c}{{\rm System~D}} \\
\hline
\ion{C}{4}  & 1 & 1.9287 & $13.71_{-0.02(-0.05)}^{+0.03(+0.11)}$ & $14.7_{-0.5(-1.6)}^{+0.5(+1.4)}$ & $1.00_{-0.04(-0.14)}^{+0.00(+0.00)}$ &                               & $0.14 \pm 0.01^f$ \\
            & 2 & 1.9290 & $12.63_{-0.18(-0.63)}^{+0.32(+2.43)}$ & (0.00 -- 12.14)$^g$        & $0.80_{-0.26(-0.65)}^{+0.20(+0.20)}$ & $1.00_{-0.03(-0.31)}^{+0.00(+0.00)}$ &                 \\
\ion{Si}{4} & 1 & 1.9287 & $12.52_{-0.12(-0.31)}^{+0.31(+1.22)}$ & $11.3_{-2.2(-6.1)}^{+2.3(+8.8)}$ & $1.00_{-0.52(-0.87)}^{+0.00(+0.00)}$ &                               & $0.02 \pm 0.01$ \\
\hline
\multicolumn{8}{c}{{\rm System~E}} \\
\hline
\ion{C}{4}  & 1 & 1.7663 & $13.47_{-0.03(-0.07)}^{+0.08(+0.28)}$ & $15.5_{-0.6(-1.9)}^{+0.7(+2.1)}$ & $1.00_{-0.16(-0.39)}^{+0.00(+0.00)}$ &                               & $0.09 \pm 0.01$ \\
            & 2 & 1.7668 & $13.10_{-0.07(-0.14)}^{+0.24(+0.70)}$ & $12.7_{-1.3(-3.7)}^{+1.5(+5.5)}$ & $0.69_{-0.15(-0.42)}^{+0.28(+0.31)}$ & $1.00_{-0.10(-0.64)}^{+0.00(+0.00)}$ & $0.04 \pm 0.01$ \\
\hline
\multicolumn{8}{c}{{\rm System~F}} \\
\hline
\ion{C}{4}  & 1 & 1.6769 & $14.00_{-0.21(-0.76)}^{+0.17(+0.43)}$ & $22.5_{-2.3(-6.9)}^{+2.5(+7.6)}$ & $0.28_{-0.04(-0.06)}^{+0.08(+0.70)}$ &                               & $0.07 \pm 0.01$ \\
\hline
\multicolumn{8}{c}{{\rm System~G}} \\
\hline
\ion{C}{4}  & 1 & 1.6387 & $13.46_{-0.04(-0.10)}^{+0.09(+0.31)}$ &  $9.8_{-0.7(-2.1)}^{+0.8(+2.3)}$ & $1.00_{-0.15(-0.38)}^{+0.00(+0.00)}$ &                               & $0.08 \pm 0.01$ 
\enddata
\tablenotetext{a}{1$\sigma$ upper/lower error bars in
  superscript/subscript. 3$\sigma$ uncertainties are in parenthesis.}
\tablenotetext{b}{Overlap covering factor ratio between components~4
  and 5 in system~C or between components~1 and 2 in systems~D and E.}
\tablenotetext{c}{Rest-frame equivalent width of blue component.}
\tablenotetext{d}{Total rest-frame equivalent widths of components~4
  and 5.}
\tablenotetext{e}{Best-fit model overestimates the depth of the red
  component, which could underestimate \cf.}
\tablenotetext{f}{Total rest-frame equivalent widths of components~1
  and 2.}
\tablenotetext{g}{This is the 3$\sigma$ distribution range since the
  1$\sigma$ distribution is multi-modal.}
\end{deluxetable}

\end{document}